\newif\ifnotblinded
\notblindedtrue  % Comment out to hide authors

\documentclass[a4paper, 11pt]{article}
\usepackage{amsmath}
\usepackage{mathtools}
\usepackage{amssymb}
\usepackage{setspace}
\usepackage{placeins}
\usepackage{minibox}
\usepackage{graphicx}
\usepackage{tikz}
\usepackage{xcolor}
\usepackage{natbib}
\usepackage{caption}
\usepackage{booktabs}
\usepackage{multirow}
\usepackage{authblk}
\usepackage{epstopdf}
\usepackage{algpseudocode}
\usepackage[hmargin = 3cm, vmargin = 2.5cm]{geometry}

\newcommand{\E}{\mathop{\mathbb E}}
\newcommand{\Var}{\mathop{\rm Var}}

\newcommand{\erf}{\mathop{\mathrm{erf}}}
\newcommand{\erfc}{\mathop{\mathrm{erfc}}}
\newcommand{\vbar}{\:\vert\:}

\newcommand{\Bvbar}{\:\Big\vert\:}
\newcommand{\ddd}{,\,\ldots,\,}

\mathtoolsset{showonlyrefs}

\begin{document}
\title{Semiparametric GARCH via Bayesian model averaging}
\ifnotblinded
\author[1]{Wilson Ye Chen}
\author[2]{Richard H. Gerlach}
\affil[1]{School of Mathematical and Physical Sciences, University of Technology Sydney}
\affil[1]{Australian Research Council Centre for Excellence in Mathematical and Statistical Frontiers}
\affil[2]{Discipline of Business Analytics, The University of Sydney}
\fi
\date{August 24, 2017}
\maketitle

\begin{abstract}
As the dynamic structure of the financial markets is subject to dramatic changes, a model capable of providing consistently accurate volatility estimates must not make strong assumptions on how prices change over time. Most volatility models impose a particular parametric functional form that relates an observed price change to a volatility forecast (news impact function). We propose a new class of functional coefficient semiparametric volatility models where the news impact function is allowed to be any smooth function, and study its ability to estimate volatilities compared to the well known parametric proposals, in both a simulation study and an empirical study with real financial data. We estimate the news impact function using a Bayesian model averaging approach, implemented via a carefully developed Markov chain Monte Carlo (MCMC) sampling algorithm. Using simulations we show that our flexible semiparametric model is able to learn the shape of the news impact function from the observed data. When applied to real financial time series, our new model suggests that the news impact functions are significantly different in shapes for different asset types, but are similar for the assets of the same type.

\vspace{0.5cm}
\end{abstract}

\begin{onehalfspace}
\section{Introduction}
\label{sec:intro}
Extensions of the GARCH model have been actively developed in the literature for the past three decades since the seminal studies by \citet{Engle1982} and \citet{Bollerslev1986}. As the normality assumption of the innovation distribution of the GARCH model is often inadequate for describing the excess kurtosis of the daily returns, one stream of studies focuses on GARCH models with flexible parametric innovation distributions. Examples include the Student's t distributions \citep{Bollerslev1987}, generalised error distribution (GED) \citep{Nelson1991}, skewed-t distribution \citep{Hansen1994}, normal inverse Gaussian (NIG) distribution \citep{JensenLunde2001}, exponential generalized beta distribution of the second kind (EGB2) \citep{WangEtAl2001}, asymmetric Laplace distribution \citep{ChenGerlachLu2012}, and the two-sided Weibull distribution \citep{ChenGerlach2013}. Another line of research focuses on developing flexible conditional volatility dynamics. Examples include the EGARCH \citep{Nelson1991}, GJR-GARCH \citep{GlostenJagannathanRunkle1993}, threshold-GARCH \citep{Zakoian1994}, NAGARCH \citep{EngleNg1993}, smooth transition GARCH \citep{Lubrano2001}, GARCH model with slowly-varying volatility component \citep{AmadoTerasvirta2013}, Beta-t-GARCH \citep{HarveyChakravarty2009}, and the more general family of score driven models \citep{CrealKoopmanLucas2013, Harvey2013}.

If either the innovation distribution or the conditional volatility dynamics is allowed to be of unknown form and estimated nonparametrically, the resulting model is referred to as a \emph{semiparametric} GARCH model. To avoid efficiency loss or inconsistency due to misspecification, several studies propose to estimate the innovation density with a nonparametric estimator, such as the kernel density estimator. For example, see \citet{EngleGonzalezRivera1991}, \citet{DrostKlaassen1997}, \citet{SunStengos2006}, and \citet{BlasquesJiLucas2015} for details. Let us introduce a few notations before proceeding with the semiparametric GARCH models of the latter kind, which is the focus of this paper.

Consider a discrete-time real-valued stochastic process denoted by $\{y_{t}\colon\, t \in \mathbb{Z}\}$ (The abbreviated notation $\{y_{t}\}$ is used hereafter.), such that $y_{t} = \sigma_{t}\epsilon_{t}$, where $\{\epsilon_{t}\}$ is an i.i.d. sequence with $\E(\epsilon_{t}) = 0$ and $\Var(\epsilon_{t}) = 1$. By construction, $\{y_{t}\}$ is a martingale difference sequence. It follows that $\sigma^{2}_{t}$ is the \emph{conditional} variance of $y_{t}$; i.e., $\Var(y_{t} \vbar \mathcal{F}_{t-1}) = \sigma^{2}_{t}$, where $\mathcal{F}_{t-1} = \sigma(\{y_{s}\colon\, s \le t-1\})$ denotes the smallest $\sigma$-algebra containing the past observations of the process, and represents the available information at $t-1$. Assume that $\sigma^{2}_{t}$ is the value of a function of possibly infinitely many past observations,
\begin{equation}
\label{eq:hfunc}
\sigma^{2}_{t} = h(y_{t-1},\, y_{t-2},\, \ldots\, \text{ad inf}),
\end{equation}
where $h(\cdot)$ is referred to as the \emph{volatility function}.

Most extensions of the GARCH model, including those with nonparametrically estimated densities, make particular parametric assumptions about the form of the volatility function $h(\cdot)$, which plays a key role in the modelling of the conditional volatility process. With the \emph{hope} of not having the need to choose amongst the various alternative parametric proposals, several studies focused on the nonparametric generalisations of the volatility function. Studies such as \citet{PaganSchwert1990} and \citet{MasryTjostheim1995} assumed that $h(\cdot)$ is a smooth but unknown $J$-dimensional function for some $J < \infty$,
\begin{equation}
\label{eq:hfunc_jdim}
\sigma^{2}_{t} = h(y_{t-s_{1}} \ddd y_{t-s_{J}}),
\end{equation}
with the lags $\{s_{1} \ddd s_{J}\}$ selected a priori, and estimated $h(\cdot)$ nonparametrically using a $J$-dimensional smoother. For example, notice that \eqref{eq:hfunc_jdim} can be written in the form
\begin{equation}
\label{eq:hfunc_jdim_reg}
y^{2}_{t} = h(y_{t-s_{1}} \ddd y_{t-s_{J}}) + u_{t},
\end{equation}
where $u_{t} = \sigma^{2}_{t}(\epsilon^{2}_{t} - 1)$ and $\{u_{t}\}$ is a martingale difference sequence, which suggests that one can estimate $h(\cdot)$ using a kernel regression method by regressing $y^{2}_{t}$ on the lagged observations $y_{t-s_{1}} \ddd y_{t-s_{J}}$. In practice, the model in \eqref{eq:hfunc_jdim} can only be estimated for a small $J$ as it suffers from the so called ``\emph{curse of dimensionality}"; the optimal rate of convergence of a consistent nonparametric estimator of $h(\cdot)$ decreases rapidly as the dimensionality $J$ increases; see for example \citet{Stone1980} and \citet{Hansen2008}. As the conditional variance processes estimated from many empirical datasets are highly persistent, a large value of $J$ is typically needed for the model in \eqref{eq:hfunc_jdim} to adequately capture the empirically observed high persistence. In order to improve the best achievable rate of convergence, studies such as \citet{Hafner1998}, \citet{CarrollHardleMammen2002}, \citet{Yang2002}, and \citet{WangEtAl2012} focused on additive models of the form
\begin{equation}
\label{eq:hfunc_add}
\sigma^{2}_{t} = \sum_{i=1}^{J}\psi_{i}(\boldsymbol{\theta})m(y_{t-i}),
\end{equation}
for some $J < \infty$, where the coefficient functions $\{\psi_{1}(\boldsymbol{\theta}) \ddd \psi_{J}(\boldsymbol{\theta})\}$ are assumed known up to a parameter vector $\boldsymbol{\theta}$, and the smooth but unknown univariate function $m(\cdot)$ is to be estimated nonparametrically. Studies such as \citet{GourierouxMonfort1992}, \citet{EngleNg1993}, and \citet{LintonMammen2005} considered a more general case where $J = \infty$ in \eqref{eq:hfunc_add}. Their models include the popular GARCH$(1,1)$ model as a special case, while \eqref{eq:hfunc_add} is a semiparametric ARCH$(J)$ model. Notice that in \eqref{eq:hfunc_add} if we let $J = \infty$, $\boldsymbol{\theta} = \beta$, and $\psi_{i}(\beta) = \beta^{i-1}$, with $\beta \in (0,1)$, then the volatility function in \eqref{eq:hfunc_add} can be written in the form $\sigma^{2}_{t} = \beta\sigma^{2}_{t-1} + m(y_{t-1})$. The function $m(\cdot)$ in \eqref{eq:hfunc_add} is often referred to as the ``\emph{news impact function}" in the literature, as it updates the conditional volatility based on the new information contained in the observation $y_{t-1}$. Another alternative strategy for modelling the volatility function $h(\cdot)$ was investigated in \citet{BuhlmannMcNeil2002} and \citet{AudrinoBuhlmann2009} where $h(\cdot)$ was assumed to be a smooth but unknown bivariate function
\begin{equation}
\label{eq:hfunc_bivar}
\sigma^{2}_{t} = h(y_{t-1},\, \sigma^{2}_{t-1}),
\end{equation}
and was estimated nonparametrically using a bivariate smoothing method.

In this paper, we propose a novel semiparametric GARCH model, where the sequence of unobserved conditional variances $\{\sigma^{2}_{t}\}$ follows a \emph{functional coefficient autoregressive} (FAR) process. We model the coefficient function by a \emph{quadratic regression spline}, where both the number and the positions of the knots are selected using a Bayesian model averaging (BMA) approach. A crucial step in BMA is to compute the expectations of random variables of interest with respect to the \emph{joint} posterior distribution of the model parameters and the knot placements. For this purpose, we develop an adaptive Markov chain Monte Carlo (MCMC) algorithm to simulate from such joint posterior distribution. Similar to the additive model in \eqref{eq:hfunc_add}, the proposed approach only requires the estimation of an unknown univariate function, however, unlike \eqref{eq:hfunc_add}, our model nests the GARCH$(1,1)$ model. Compared with a kernel regression method with a global bandwidth, a regression spline with strategically placed knots has the advantage of being locally adaptive \citep{SmithKohn1996}, and thus can accommodate varying degrees of smoothness across regions of the domain.

The rest of the paper is structured as follows. Section \ref{sec:model} presents our semiparametric GARCH model, shows how various parametric models can be written as special cases of the proposed model, and defines the spline functional coefficient. Section \ref{sec:bayesian} describes the model in a Bayesian context, and explains how the scalar parameters and the functional coefficient are estimated using a MCMC sampling algorithm. Section \ref{sec:simulation} presents a simulation study comparing the performance of the proposed model to the popular parametric alternatives. Section \ref{sec:empirical} demonstrates the proposed model by applying it to ten real financial time series, and presents a model comparison study based on the averaged DIC -- a slight generalisation of the usual single model DIC.

\section{The proposed model}
\label{sec:model}
\subsection{Functional coefficient semiparametric GARCH}
\label{sec:funcoef}
Continuing with the notations introduced in Section~\ref{sec:intro}, let $\{r_{t}\colon\, t \in \mathbb{Z}\}$ denote a real-valued discrete-time stochastic process, and $\mathcal{F}_{t} = \sigma(\{r_{s}\colon\, s \le t\})$ denote the natural filtration of $\{r_{t}\}$. Our proposed functional coefficient semiparametric GARCH (denoted by SP-GARCH hereafter) model is defined as follows.
\begin{equation}
\label{eq:spgarch}
\begin{aligned}
r_{t} &= \mu + y_{t}, \\
y_{t} &= \sigma_{t}\epsilon_{t}, \\
\sigma^{2}_{t} &= \omega + g(\epsilon_{t-1})\sigma^{2}_{t-1}, \\
\end{aligned}
\end{equation}
where $\{\epsilon_{t}\}$ is an i.i.d. sequence of innovations with $\E(\epsilon_{t}) = 0$ and $\Var(\epsilon_{t}) = 1$. It follows that $\mu$ and $\sigma^{2}_{t}$ are, respectively, the conditional mean and conditional variance of $r_{t}$; i.e., $\E(r_{t} \vbar \mathcal{F}_{t-1}) = \mu$ and $\Var(r_{t} \vbar \mathcal{F}_{t-1}) = \sigma^{2}_{t}$. In \eqref{eq:spgarch}, we make the assumption that the conditional mean of $r_{t}$ is constant over time. The constant mean assumption is often reasonable in empirical applications such as when $\{r_{t}\}$ is a model for a sequence of daily returns of a financial asset \citep{Taylor2011}. The distribution of the innovation $\epsilon_{t}$ can either be chosen from a wide range of parametric families with finite variance such as those mentioned in Section~\ref{sec:intro} or modelled nonparametrically. In this paper, we take a parametric approach and assume that $\epsilon_{t} \sim F_{\mathrm{sdt},\nu}$, for $\nu > 2$, where $F_{\mathrm{sdt},\nu}$ denotes a \emph{standardised} Student-t distribution with $\nu$ degrees of freedom. (A random variable $z$ follows $F_{\mathrm{sdt},\nu}$ if it is given by $z = x[(\nu - 2) / \nu]^{1/2}$, where $x$ follows a Student-t distribution with $\nu$ degrees of freedom. It follows that $\E(z) = 0$ and $\Var(z) = 1$.) The typical assumption that $\epsilon_{t}$ follows a heavy-tailed distribution is motivated by the stylised fact that the distribution of daily returns, standardised using an estimated volatility model, is heavy-tailed and approximately symmetric for many financial assets \citep{Taylor2011}.

The proposed model in \eqref{eq:spgarch} admits the interpretation that the sequence of unobserved conditional variances $\{\sigma^{2}_{t}\}$ follows an autoregressive process with a functional coefficient $g(\cdot)$. It can be seen that the function $g(\cdot)$ plays a similar role to that of the news impact function $m(\cdot)$ in \eqref{eq:hfunc_add}, as it determines how $\sigma^{2}_{t-1}$ is updated given the value of the innovation $\epsilon_{t-1}$, which can be considered as news. Notice that the argument of $g(\cdot)$ is the lagged innovation $\epsilon_{t-1}$, as opposed to the lagged centred observation $y_{t-1}$. We consider this as an advantage of the new model over the previous semiparametric proposals. For example, from a modeller's perspective, since the distribution of $\epsilon_{t}$ is either fully known or at least known up to its third moment (by construction), and is usually not time dependent, one can utilise such prior knowledge to improve the statistical properties of an estimator by imposing a certain structure on $g(\cdot)$. Furthermore, because the scale of $\epsilon_{t}$ is the same across different datasets, the estimates of $g(\cdot)$ can be directly compared in empirical applications.

Notice that $g(\epsilon_{t})$ is an i.i.d. random variable. Let us suppose that $g(\epsilon_{t})$ is replaced by a constant $\gamma$ and the process $\{\sigma^{2}_{t}\}$ has an initial value of $\sigma^{2}_{0} > 0$, so that $\{\sigma^{2}_{t}\}$ becomes a fully deterministic process. For $\gamma \in (0,1)$, we can see that $\gamma$ controls the rate at which $\sigma^{2}_{t}$ approaches its lower bound $\omega / (1 - \gamma)$ as $t \to \infty$. Since $g(\epsilon_{t})$ is a random variable, the quantity $\E[g(\epsilon_{t})]$ controls the \emph{average} rate of decay of $\{\sigma^{2}_{t}\}$, and is referred to as the \emph{persistence} of the conditional variance process. The process $\{r_{t}\}$ defined in \eqref{eq:spgarch} is \emph{covariance stationary} if
\begin{equation}
\label{eq:stationarity}
\E[g(\epsilon_{t})] \in [0, 1).
\end{equation}
If $\{r_{t}\}$ is covariance stationary, its \emph{unconditional} variance is given by
\begin{equation}
\label{eq:uncvar}
\Var(r_{t}) = \E(\sigma_{t}^{2}) = \frac{\omega}{1 - \E[g(\epsilon_{t})]}.
\end{equation}

\begin{figure}[h]
\centering
\centerline{\includegraphics[width = 0.9 \textwidth]{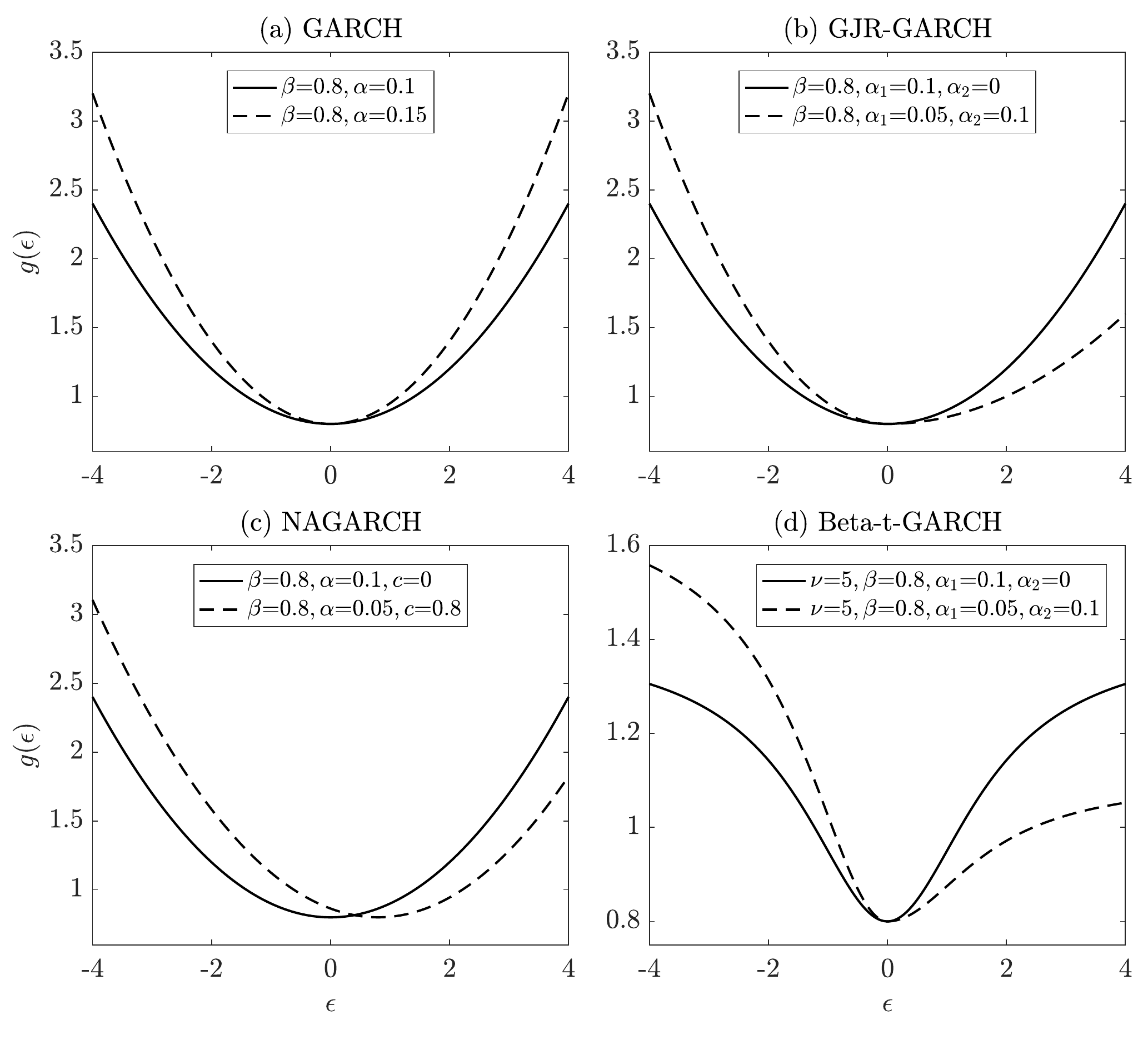}}
\caption{Plots of $g(\cdot)$ in $[-4, 4]$ for various parametric GARCH models.}
\label{fig:gfunc_para}
\end{figure}

The GARCH$(1,1)$ model and many of its popular parametric extensions can be written as special cases of the new SP-GARCH model in \eqref{eq:spgarch} with particular forms of the coefficient function $g(\cdot)$. For example, Table~\ref{tab:gfunc_para} lists the forms of the coefficient function when the first order GARCH, GJR-GARCH, NAGARCH, and Beta-t-GARCH models are written as special cases of \eqref{eq:spgarch}, where $\beta$, $\alpha$, $\alpha_{1}$, $\alpha_{2}$, $c$, and $\nu$ are parameters, taking their values typically in a subset of $\mathbb{R}$, and $I_{(-\infty,0)}(\epsilon)$ takes the value 1 if $\epsilon < 0$ and 0 otherwise. Figure~\ref{fig:gfunc_para} shows the corresponding plots of these functions in the interval $[-4, 4]$ and how their shapes change for the various values of their parameters.

Notice that all the models except for the GARCH allow a negative innovation to contribute differently to the conditional variance from a positive innovation of the same magnitude. For the GJR-GARCH and the Beta-t-GARCH models, the asymmetry of $g(\cdot)$ is created by scaling an even function differently on each side of zero. The scaling factor is $\alpha_{1} + \alpha_{2}$ for a negative argument, and is $\alpha_{1}$ for a positive argument. Moreover, the NAGARCH model is the only one that allows the minimum of $g(\cdot)$ to occur away from 0. For the Beta-t-GARCH model, the shape of $g(\cdot)$ is directly linked to the innovation distribution, which is assumed to be $F_{\mathrm{sdt},\nu}$.

The Beta-t-GARCH model can be considered as a specific member of the family of \emph{score driven} models known as generalised autoregressive score (GAS) models in \citet{CrealKoopmanLucas2013} and dynamic conditional score (DCS) models in \citet{Harvey2013}. In a score driven volatility model, $g(\cdot)$ is a transformed gradient function (i.e., score) of the conditional distribution of $r_{t}$ with respect to $\sigma_{t}^{2}$. The gradient function of the Student-t distribution dampens the effect of extreme observations on conditional variance when the degree of freedom parameter is small. Notice that $u \to \epsilon^{2}$ as $\nu \to \infty$. Hence, the Beta-t-GARCH model is reduced to GJR-GARCH when the innovation distribution is Gaussian.

It is straightforward to show that the stationarity condition and the unconditional variance of each of the parametric models is consistent with those of the new model in \eqref{eq:stationarity} and \eqref{eq:uncvar}.

\begin{table}[h]
\centering
\begin{tabular}{ll}
\toprule
GARCH & $g(\epsilon) = \beta + \alpha \epsilon^{2}.$ \\
\midrule
GJR-GARCH & $g(\epsilon) = \beta + [\alpha_{1} + \alpha_{2} I_{(-\infty,0)}(\epsilon)] \epsilon^{2}.$ \\
\midrule
NAGARCH & $g(\epsilon) = \beta + \alpha (\epsilon - c)^{2}.$ \\
\midrule
Beta-t-GARCH & $g(\epsilon) = \beta + [\alpha_{1} + \alpha_{2} I_{(-\infty,0)}(\epsilon)] u,$ \\
& $u = (\nu + 1) \epsilon^{2} / (\nu - 2 + \epsilon^{2}).$ \\
\bottomrule
\end{tabular}
\caption{Examples of specific functional forms of $g(\cdot)$ in \eqref{eq:spgarch}.}
\label{tab:gfunc_para}
\end{table}
\FloatBarrier

\FloatBarrier

\subsection{Spline coefficient function}
\label{sec:spline}
In this paper, we generalise the various parametric proposals by assuming that the coefficient function $g(\cdot)$ in \eqref{eq:spgarch} is an unknown but smooth univariate function. We model $g(\cdot)$ nonparametrically by a polynomial regression spline. A \emph{$p$th degree polynomial spline}, for $p \in \mathbb{N}_{0} = \{0, \mathbb{N}\}$, is a function constructed piece-wise where each piece is a $p$th degree polynomial function. The set of points in the domain at which the pieces are joint together are referred to as \emph{knots}. It is worth mentioning at this point that we consider a so called \emph{regression spline} approach where there are significantly fewer number of knots than the number of observations, and the knot positions are selected as part of the modelling task. For example, a simple strategy is to place the knots either equally spaced or at specific sample quantiles. This is different from the \emph{smoothing spline} approach (See \citet{Wahba1990}, for example.), where the knots are the observations themselves and a roughness penalty is added explicitly to reduce the effective degrees of freedom.

Specifically, the coefficient function in our model is given by
\begin{equation}
\label{eq:spline_p}
g(\epsilon) = \sum_{i=0}^{p} b_{i} \epsilon^{i} + \sum_{i=1}^{K}\beta_{i} (\epsilon - k_{i})^{p}_{+},
\end{equation}
where $p \in \mathbb{N}_{0}$ is the degree of the spline, $K \in \mathbb{N}_{0}$ is the number of knots, $\{k_{1} \ddd k_{K}\}$ is the sequence of knots satisfying $k_{1} < \cdots < k_{K}$, and $(\epsilon - k_{i})^{p}_{+}$ is known as a $p$th degree \emph{truncated power function} defined by
\begin{equation*}
(\epsilon - k_{i})^{p}_{+} = (\epsilon - k_{i})^{p}I_{[k_{i},\infty)}(\epsilon),
\end{equation*}
with $I_{[k_{i},\infty)}(\epsilon)$ taking the value 1 if $\epsilon \ge k_{i}$ and 0 otherwise. In \eqref{eq:spline_p}, the $p$th degree spline is represented as a linear combination of the functions
\begin{equation}
\label{eq:spline_basis}
\epsilon^{0} \ddd \epsilon^{p},\, (\epsilon - k_{1})^{p}_{+} \ddd (\epsilon - k_{K})^{p}_{+},
\end{equation}
where $b_{0} \ddd b_{p},\, \beta_{1} \ddd \beta_{K}$ are the coefficients in $\mathbb{R}$. It can be shown that the set of functions in \eqref{eq:spline_basis} is a basis for the space of $p$th degree splines with knots $k_{1} \ddd k_{K}$ (See \citet{DeBoor1978} and \citet{RuppertWandCarroll2003}, for example.). Furthermore, any linear combination of the functions in \eqref{eq:spline_basis} has $p - 1$ continuous derivatives. The particular basis whose members are given by \eqref{eq:spline_basis} is known as a \emph{truncated power basis}. In this paper, we restrict our attention to \emph{quadratic splines} where $p = 2$ and
\begin{equation}
\label{eq:spline_2}
g(\epsilon) = b_{0} + b_{1}\epsilon + b_{2}\epsilon^{2} + \sum_{i=1}^{K}\beta_{i} (\epsilon - k_{i})^{2}_{+}.
\end{equation}
It can be shown that, when $p = 2$, several well known parametric models can be expressed exactly in terms of the spline coefficients. For example, following from the definitions in Table~\ref{tab:gfunc_para}, the first order GARCH corresponds to the case where $b_{1} = 0$ and $K = 0$; GJR-GARCH corresponds to the case where $b_{0} = \beta$, $b_{1} = 0$, $b_{2} = \alpha_{1} + \alpha_{2}$, $\beta_{1} = -\alpha_{2}$, $K = 1$, and $k_{1} = 0$; NAGARCH corresponds to the case where $b_{0} = \beta + \alpha c^{2}$, $b_{1} = -2 \alpha c$, $b_{2} = \alpha$, and $K = 0$.

\begin{figure}[h]
\centering
\centerline{\includegraphics[width = 0.6 \textwidth]{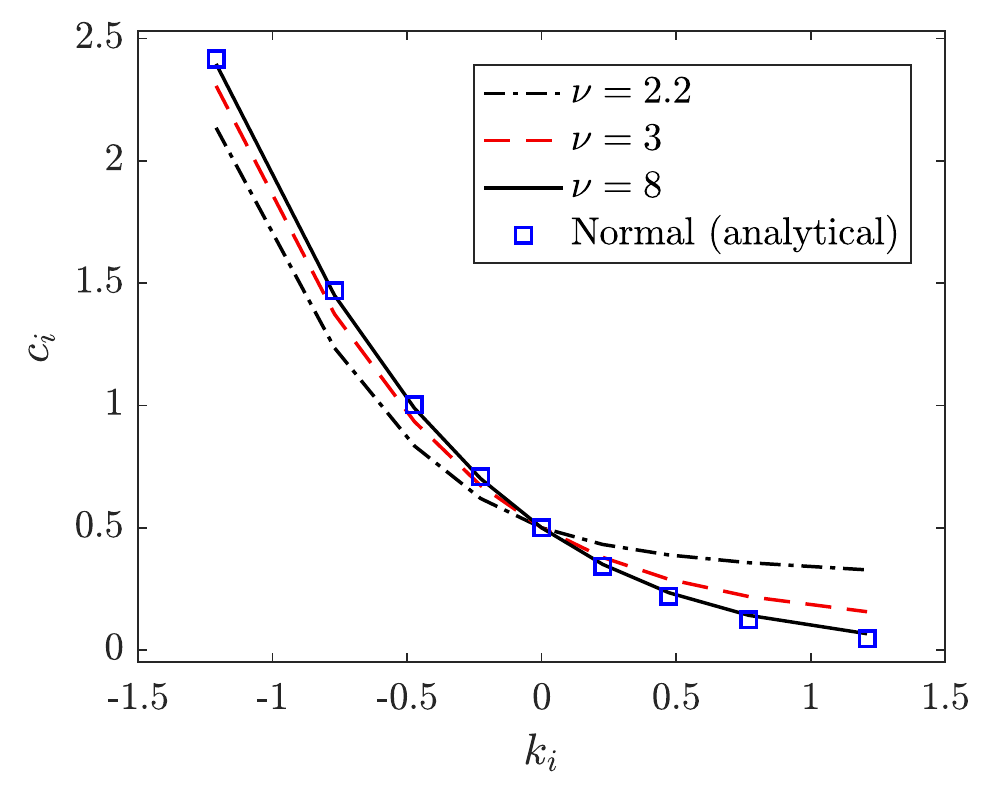}}
\caption{Values of $\{c_{i}\}$ computed using adaptive quadrature for standardised t with $\nu \in \{2.2,\, 3,\, 8\}$, and computed analytically for the standard normal.}
\label{fig:cval_t}
\end{figure}

Assume that the distribution of $\epsilon_{t}$ is both standardised and symmetric. Using the definition in \eqref{eq:spline_2} and the distributional assumption of $\epsilon_{t}$, we can derive the persistence of the SP-GARCH conditional variance process as follows.
\begin{equation}
\label{eq:ex_g}
\E[g(\epsilon_{t})] = \int_{-\infty}^{\infty}g(\epsilon_{t})f(\epsilon_{t})d\epsilon_{t} = b_{0} + b_{2} + \sum_{i=1}^{K} \beta_{i} c_{i},
\end{equation}
where $f(\cdot)$ is the density function of the innovation distribution, and $c_{i}$ is given by
\begin{equation}
\label{eq:cval}
c_{i} = \int_{k_{i}}^{\infty} (\epsilon_{t} - k_{i})^{2} f(\epsilon_{t}) d \epsilon_{t}.
\end{equation}
The integral for $c_{i}$ in \eqref{eq:cval} can be obtained in closed form if $f(\cdot)$ is the density function of the standard normal distribution.
\begin{equation}
\label{eq:cval_norm}
c_{i} = - \frac{1}{\sqrt{2 \pi}} \exp\left(-\frac{k_{i}^{2}}{2}\right) k + \frac{1}{2} (1 + k_{i}^{2}) \erfc \left( \frac{k_{i}}{\sqrt{2}} \right),
\end{equation}
where $\erfc(x) = 1 - \erf(x)$ denotes the complementary error function. If $f(\cdot)$ is instead the density function of $F_{\mathrm{sdt},\nu}$, as per our assumption in Section~\ref{sec:funcoef}, the expression for $c_{i}$ is rather complicated, and the evaluation of such expression is subject to numerical difficulties. In such case, as a practical alternative, the value of $c_{i}$ can be approximated with high accuracy using a numerical integration algorithm, such as the \emph{univariate adaptive quadrature}. Figure~\ref{fig:cval_t} shows plots of $c_{1} \ddd c_{9}$ computed using the adaptive quadrature for $F_{\mathrm{sdt},\nu}$ with $\nu \in \{2.2,\, 3,\, 8\}$, and using the closed-form expression in \eqref{eq:cval_norm} for the standard normal distribution, where the corresponding knots $k_{1} \ddd k_{9}$ are placed at the quantile levels $0.1 \ddd 0.9$ of $F_{\mathrm{sdt},8}$. It can be seen that, even at $\nu = 8$, the values of $\{c_{i}\}$ are already very close to those computed analytically for the standard normal distribution, which suggests that, for the t distribution, equation \eqref{eq:cval_norm} can be used to approximate the integral for $c_{i}$ when $\nu$ is moderate or large (for example, $\nu \ge 8$).

\FloatBarrier

\section{Bayesian inference}
\label{sec:bayesian}
\subsection{Bayesian model averaging}
\label{sec:bma}
Our goal in Section~\ref{sec:bayesian} is to estimate the degrees of freedom $\nu$, the mean $\mu$, the constant parameter $\omega$ of the volatility function, and the coefficient function $g(\cdot)$ of the SP-GARCH model defined in \eqref{eq:spgarch}, where $g(\cdot)$ is modelled as a quadratic regression spline with a sequence of knots $\{k_{1} \ddd k_{K}\}$ defined in \eqref{eq:spline_2}. The choice of the knot sequence is important in estimating functions by regression splines, as it imposes a certain structure on the estimates. Ideally, we would want a small number of well positioned knots, so that we have a parsimonious model that can also account for the important details in the data. We employ a Bayesian variable selection approach where \emph{model averaging} occurs naturally, and in doing so, we account for two types of uncertainty: the uncertainty in the knot sequence $\{k_{1} \ddd k_{K}\}$ and the uncertainty in the parameter values conditional on a particular knot sequence. For a detailed treatment on Bayesian model selection and model averaging, see \citet{RafteryZheng2003}, \citet{ClydeGeorge2004}, and the references therein.

Suppose that there are $M \in \mathbb{N}$ candidate models, indexed by $\tau \in \{1 \ddd M\}$. Let $\boldsymbol{\theta}_{\tau} \in \Theta_{\tau}$ denote the parameter vector associated with the $\tau$th model, where the set $\Theta_{\tau}$ defines the parameter space, and let $\mathbf{r} = (r_{1} \ddd r_{T})$ denote a vector of observations. Further suppose that we are interested in estimating a quantity $\phi$, such that conditional on $\mathbf{r}$, $\phi = \eta(\boldsymbol{\theta}_{\tau},\,\tau)$, for some function $\eta(\cdot)$. For example, $\phi$ may be the one-day-ahead forecast of the conditional volatility, or the value of $g(\epsilon_{t})$ at some specific value of $\epsilon_{t}$. Under the assumption of model uncertainty, the posterior mean of $\phi$ can be taken as a Bayesian point estimate, which is obtained by averaging over all the candidate models;
\begin{equation}
\label{eq:phi_bayes}
\begin{aligned}
\hat{\phi}_{\mathrm{Bayes}} = \E(\phi \vbar \mathbf{r}) &= \sum_{\tau=1}^{M} \left[\int_{\Theta_{\tau}}\eta(\boldsymbol{\theta}_{\tau},\,\tau)p(\boldsymbol{\theta}_{\tau},\, \tau \vbar \mathbf{r})d\boldsymbol{\theta}_{\tau}\right] \\
&= \sum_{\tau=1}^{M} \left[\int_{\Theta_{\tau}}\eta(\boldsymbol{\theta}_{\tau},\,\tau)p(\boldsymbol{\theta}_{\tau} \vbar \tau,\, \mathbf{r})d\boldsymbol{\theta}_{\tau}\right]p(\tau \vbar \mathbf{r}) \\
&= \sum_{\tau=1}^{M}\E(\phi \vbar \tau,\, \mathbf{r})p(\tau \vbar \mathbf{r}).
\end{aligned}
\end{equation}
While for some problems one can obtain the closed-form expressions for both the conditional expectation $\E(\phi \vbar \tau,\, \mathbf{r})$ and the posterior model probabilities $p(\tau \vbar \mathbf{r})$ (For example, see \citet{SmithKohn1996}.), for many realistic models they are analytically intractable, and approximation methods must be used. Even in the case of \citet{SmithKohn1996}, the summation in \eqref{eq:phi_bayes} is infeasible to compute due to a large number of candidate models. We develop an adaptive MCMC algorithm to simulate from the joint posterior distribution $p(\boldsymbol{\theta}_{\tau},\, \tau \vbar \mathbf{r})$ defined on the space
\begin{equation*}
\bigcup_{\tau \in \{1 \ddd M\}} \Theta_{\tau} \times \{\tau\}.
\end{equation*}
If $\{(\boldsymbol{\theta}_{\tau}^{[i]},\, \tau^{[i]})\colon\, i \in \{1 \ddd N\}\}$ denotes a sample from $p(\boldsymbol{\theta}_{\tau},\, \tau \vbar \mathbf{r})$ generated by a MCMC sampling algorithm, an approximation of $\hat{\phi}_{\mathrm{Bayes}}$ can then be computed as
\begin{equation}
\label{eq:phi_mcmc}
\hat{\phi}_{\mathrm{MCMC}} = \frac{1}{N}\sum_{i=1}^{N}\eta(\boldsymbol{\theta}_{\tau}^{[i]},\, \tau^{[i]}).
\end{equation}
It can be seen from \eqref{eq:phi_mcmc} that model averaging is achieved without additional efforts via a single run of the MCMC sampler.

An advantage of the basis representation in \eqref{eq:spline_p} is that candidate models can be represented as \emph{submodels} (or restricted models) nested within a \emph{supermodel}. If we consider a model with a large number of knots (i.e., a large $K$) as a supermodel, then a candidate model with a specific knot sequence can be considered as a submodel with a set of knot coefficients $\{\beta_{i}\}$ restricted to zero for some $i \in \{1 \ddd K\}$. In a Bayesian approach, the knot selection can be achieved by allowing a positive prior probability that each knot coefficient $\beta_{i}$ is \emph{exactly zero}.

\subsection{Joint posterior}
\label{sec:joint_post}
In this section, we define the the joint posterior distribution of the model parameters and the knot sequences. We adopt the paradigm where there is a large number of potential knots, and a subset of which is assigned to a candidate model. Let $\boldsymbol{\theta} = (\nu,\, \mu,\, \omega,\, b_{0},\, b_{1},\, b_{2},\, \beta_{1} \ddd \beta_{K})$ denote the parameter vector of the \emph{full model}, and $\mathbf{k} = (k_{1} \ddd k_{K})$ denote the vector of the potential knots. Each candidate model is indexed by a binary vector $\mathbf{m} = (m_{1} \ddd m_{K})$ such that
\begin{equation}
\label{eq:m_vec}
m_{i} =
\begin{dcases}
0 & \text{if $\beta_{i} = 0$,} \\
1 & \text{if $\beta_{i} \ne 0$.}
\end{dcases}
\end{equation}
The joint posterior distribution of the pair $(\boldsymbol{\theta},\, \mathbf{m})$ given data is then proportional to the likelihood multiplied by the joint prior;
\begin{equation}
\label{eq:posterior}
p(\boldsymbol{\theta},\, \mathbf{m} \vbar \mathbf{r}) \propto p(\mathbf{r} \vbar \boldsymbol{\theta},\, \mathbf{m}) p(\boldsymbol{\theta},\, \mathbf{m}).
\end{equation}

Following from the model definition in \eqref{eq:spgarch} and \eqref{eq:spline_2}, the likelihood is given by
\begin{equation}
\label{eq:likelihood}
p(\mathbf{r} \vbar \boldsymbol{\theta},\, \mathbf{m}) = p(\mathbf{r} \vbar \boldsymbol{\theta}) = I_{\Theta}(\boldsymbol{\theta}) \prod_{t=1}^{T}p(r_{t} \vbar \mathcal{F}_{t-1},\, \boldsymbol{\theta}),
\end{equation} 
where the conditional distribution of each observation $r_{t}$ is a Student-t distribution with $\nu$ degrees of freedom, whose mean is $\mu$ and whose variance is $\sigma^{2}_{t}$, and $I_{\Theta}(\boldsymbol{\theta})$ is equal to one if $\boldsymbol{\theta} \in \Theta$ and zero otherwise. It should be made clear in \eqref{eq:likelihood} that $\mathbf{r}$ and $\mathbf{m}$ are independent conditional on $\boldsymbol{\theta}$; the vector $\boldsymbol{\theta}$ encodes all the information needed to generate $\mathbf{r}$, as a result of the nested model set-up. We assume that $\mathcal{F}_{0} = \sigma(\varnothing)$, and set the initial conditional variance $\sigma^{2}_{1}$ to the sample variance of the realisation of $\mathbf{r}$. We use the indicator function $I_{\Theta}(\cdot)$ to restrict the parameter space, by setting the likelihood to zero when $\boldsymbol{\theta} \notin \Theta$. The restricted parameter space $\Theta$ is defined as
\begin{equation}
\label{eq:par_space}
\Theta = \left\{\boldsymbol{\theta}\colon\;
\minibox{
$2 < \nu \le 200$, \\
$\omega > 0$, \\
$0 < b_{0} + b_{2} + \sum_{i=1}^{K}\beta_{i}c_{i} < 1$, \\
$\sigma^{2}_{t} > 0$, for $t \in \{1 \ddd T+1\}$
}
\right\},
\end{equation}
so that when $\boldsymbol{\theta} \in \Theta$, the process $\{r_{t}\}$ is covariance stationary with a well defined unconditional variance given by \eqref{eq:uncvar}, and the process $\{\sigma^{2}_{t}\}$ is positive for $t \in \{1 \ddd T+1\}$. Notice that the positivity restriction on the conditional variance process in \eqref{eq:par_space} is data-dependent, as $\sigma^{2}_{t}$ is a function of both $\boldsymbol{\theta}$ and $\mathbf{r}$. This type of \emph{implicit} restrictions was used in \citet{HudsonGerlach2008} to replace the sufficient conditions, which are explicitly available in terms of the model parameters, but more restrictive. The constants $\{c_{i}\}$ are given by \eqref{eq:cval}. Since the density function $f(\cdot)$ in \eqref{eq:cval} is of $F_{\mathrm{sdt},\nu}$, the integral is evaluated numerically using adaptive quadrature. From an implementation perspective, these constants can be computed ex-ante on a fine grid for a range of degrees of freedom $\nu$, and stored in a lookup table. During the evaluation of the indicator function $I_{\{\boldsymbol{\theta}\in\Theta\}}(\boldsymbol{\theta})$, a table-lookup operation is computationally less expensive than running a quadrature algorithm.

The joint prior can be constructed as
\begin{equation}
\label{eq:prior_joint}
p(\boldsymbol{\theta},\, \mathbf{m}) = p(\boldsymbol{\theta} \vbar \mathbf{m}) p(\mathbf{m}).
\end{equation}
When $\mathbf{m}$ is known, we define the conditional prior for $\boldsymbol{\theta}$ as
\begin{equation}
\label{eq:prior_theta}
p(\boldsymbol{\theta} \vbar \mathbf{m}) \propto \nu^{-2}\prod_{i=1}^{K} \left[ m_{i}f_{\mathrm{Gauss}}(\beta_{i};\; 0,\, \sigma^{2}_{\beta}) + (1 - m_{i})\delta(\beta_{i})\right],
\end{equation}
where $f_{\mathrm{Gauss}}(\beta_{i};\; 0,\, \sigma^{2}_{\beta})$ is a Gaussian density, with mean zero and variance $\sigma^{2}_{\beta}$, the delta function $\delta(\beta_{i})$ represents a \emph{point probability mass} at $\beta_{i} = 0$. Conditional on $\mathbf{m}$, the parameters are assumed to be independent a priori. The conditional prior in \eqref{eq:prior_theta} implies that $\beta_{i}$ is known to be exactly zero a priori when $m_{i} = 0$. As the prior is centred at zero for each $\beta_{i}$, the value of $\sigma^{2}_{\beta}$ controls the level of shrinkage of the nonzero knot coefficients. We treat $\sigma^{2}_{\beta}$ as a tuning parameter whose value is set by the user. From \eqref{eq:prior_theta}, we can see that the prior for $\nu$ behaves similarly to a half-Cauchy prior with a scale parameter equal to one. It can be shown that this prior is flat on $\nu^{-1}$. Similar priors for $\nu$ were employed in studies such as \citet{BauwensLubrano1998} and \citet{ChenEtAl2012} in the context of the GARCH models. For all the parameters other than $\nu$ and $\beta_{1} \ddd \beta_{K}$, the conditional prior in \eqref{eq:prior_theta} is flat. Finally, a flat prior is used for the model probabilities,
\begin{equation}
\label{eq:prior_m}
p(\mathbf{m}) = 0.5^{K} \propto 1,
\end{equation}
so that each model is given an equal prior probability. More generally, one can use the independent Bernoulli priors, $p(\mathbf{m}) = \prod_{i=1}^{K}[\pi_{i}^{m_{i}}(1 - \pi_{i})^{(1 - m_{i})}]$, with different values of $\pi_{1} \ddd \pi_{K}$ to express a prior preference for certain knots over the others. The main feature of the the joint prior specified in \eqref{eq:prior_joint} to \eqref{eq:prior_m} is that each knot coefficient $\beta_{i}$ is given a positive probability a priori to be exactly zero. This type of prior for $\beta_{i}$ is known in the literature as the spike-and-slab prior \citep{MitchellBeauchamp1988, GeorgeMcCulloch1993}.

\subsection{Adaptive MCMC algorithm}
\label{sec:mcmc}
As the closed-form expressions are unavailable for the exact Bayesian estimates of the type in \eqref{eq:phi_bayes} based on the joint-posterior distribution of our model, we develop an adaptive MCMC sampling algorithm to simulate from $p(\boldsymbol{\theta},\, \mathbf{m} \vbar \mathbf{r})$ defined in \eqref{eq:posterior} to \eqref{eq:prior_m}, and compute the approximate estimates using \eqref{eq:phi_mcmc}. The aim is to construct an ergodic Markov chain whose stationary distribution is $p(\boldsymbol{\theta},\, \mathbf{m} \vbar \mathbf{r})$, and then generate a sample path $\{(\boldsymbol{\theta}^{[i]},\, \mathbf{m}^{[i]})\colon\, i \in \{1 \ddd N\}\}$ of the chain. In general, our sampling algorithm involves the following steps:
\begin{enumerate}
\item Let $\mathbf{m}^{[1]}$ be an arbitrary binary vector of length $K$, and $i = 1$.
\item Generate a proposal $(\boldsymbol{\theta}^{*},\, \mathbf{m}^{*})$ from the joint distribution $q(\boldsymbol{\theta}^{*},\, \mathbf{m}^{*} \vbar \mathbf{m}^{[i]})$ by
\begin{enumerate}
\item generating $\mathbf{m}^{*}$ from $q(\mathbf{m}^{*} \vbar \mathbf{m}^{[i]})$,
\item generating $\boldsymbol{\theta}^{*}$ from $q(\boldsymbol{\theta}^{*} \vbar \mathbf{m}^{*},\, \mathbf{m}^{[i]})$.
\end{enumerate}
\item Let $(\boldsymbol{\theta}^{[i+1]},\, \mathbf{m}^{[i+1]}) = (\boldsymbol{\theta}^{*},\, \mathbf{m}^{*})$ with probability $u$, or $(\boldsymbol{\theta}^{[i+1]},\, \mathbf{m}^{[i+1]}) = (\boldsymbol{\theta}^{[i]},\, \mathbf{m}^{[i]})$ with probability $1 - u$.
\item Increment $i$ by one, and repeat from step 2 until $i = N$.
\end{enumerate}

In step 2, the joint proposal distribution is constructed in a similar way to that of the prior distribution;
\begin{equation}
\label{eq:joint_proposal}
q(\boldsymbol{\theta}^{*},\, \mathbf{m}^{*} \vbar \mathbf{m}^{[i]}) = q(\boldsymbol{\theta}^{*} \vbar \mathbf{m}^{*},\, \mathbf{m}^{[i]})q(\mathbf{m}^{*} \vbar \mathbf{m}^{[i]}),
\end{equation}
where $\boldsymbol{\theta}^{*} = (\nu^{*},\, \mu^{*},\, \omega^{*},\, b_{0}^{*},\, b_{1}^{*},\, b_{2}^{*},\, \beta_{1}^{*} \ddd \beta_{K}^{*})$ and $\mathbf{m}^{*} = (m_{1}^{*} \ddd m_{K}^{*})$. We can generate a realisation $(\boldsymbol{\theta}^{*},\, \mathbf{m}^{*})$ from \eqref{eq:joint_proposal} by first generating a knot configuration $\mathbf{m}^{*}$, followed by generating a parameter vector $\boldsymbol{\theta}^{*}$ conditional on $\mathbf{m}^{*}$, as in the steps 2(a) and 2(b). In step 2(a), conditional on $\mathbf{m}^{[i]}$, we generate a knot configuration $\mathbf{m}^{*}$ from the following Bernoulli mixture proposal density.
\begin{equation}
\label{eq:m_proposal}
q(\mathbf{m}^{*} \vbar \mathbf{m}^{[i]}) = \prod_{j=1}^{K}\left[m^{[i]}_{j}f_{\mathrm{Bern}}(m^{*}_{j};\; 1-\pi) + (1-m^{[i]}_{j})f_{\mathrm{Bern}}(m^{*}_{j};\; \pi)\right],
\end{equation}
where $f_{\mathrm{Bern}}(m^{*}_{j};\; \pi) = \pi^{m^{*}_{j}}(1-\pi)^{(1-m^{*}_{j})}$. By employing the proposal density in \eqref{eq:m_proposal}, the resulting chain $\{\mathbf{m}^{[i]}\}$ is a random walk in the $K$-dimensional binary vector space, where the parameter $\pi$ controls the step-size. When $\pi < 0.5$, the realisations of $m^{[i]}_{j}$ and $m^{[i+1]}_{j}$ are more likely to be the same, and when $\pi > 0.5$, the opposite is true.

Let $\boldsymbol{\theta}_{\mathbf{m}}^{*}$ denote a $(6 + \sum_{i=1}^{K}m_{i}^{*})$-dimensional vector containing only those elements of $\boldsymbol{\theta}^{*}$ that are known a priori to be nonzero (I.e., $\boldsymbol{\theta}_{\mathbf{m}}^{*}$ is the parameter vector of the submodel implied by $\mathbf{m}^{*}$), and $\boldsymbol{\theta}_{\neg\mathbf{m}}^{*}$ denote a $(\sum_{i=1}^{K}(1 - m_{i}^{*}))$-dimensional vector containing only those knot coefficients that correspond to the zero elements of $\mathbf{m}^{*}$. In step 2(b), when $\mathbf{m}^{*}$ is known, we write the proposal density for $\boldsymbol{\theta}^{*}$ as
\begin{equation}
\label{eq:full_proposal}
q(\boldsymbol{\theta}^{*} \vbar \mathbf{m}^{*},\, \mathbf{m}^{[i]}) = q(\boldsymbol{\theta}^{*} \vbar \mathbf{m}^{*}) = q(\boldsymbol{\theta}_{\mathbf{m}}^{*} \vbar \mathbf{m}^{*})q(\boldsymbol{\theta}_{\neg\mathbf{m}}^{*} \vbar \mathbf{m}^{*}),
\end{equation}
where $q(\boldsymbol{\theta}_{\neg\mathbf{m}}^{*} \vbar \mathbf{m}^{*}) = \delta(\boldsymbol{\theta}_{\neg\mathbf{m}}^{*})$ is a multivariate delta function representing a point mass at $\boldsymbol{\theta}_{\neg\mathbf{m}}^{*} = (0 \ddd 0)$. Notice that $\boldsymbol{\theta}^{*}$ and $\mathbf{m}^{[i]}$ are conditionally independent given $\mathbf{m}^{*}$. We generate $\boldsymbol{\theta}_{\mathbf{m}}^{*}$ from a mixture of multivariate Gaussian distributions with a different scale for each component,
\begin{equation}
\label{eq:sub_proposal}
q(\boldsymbol{\theta}_{\mathbf{m}}^{*} \vbar \mathbf{m}^{*}) = \sum_{j=1}^{N_{\mathrm{mix}}} w_{j} f_{\mathrm{mvn}}(\boldsymbol{\theta}_{\mathbf{m}}^{*};\; \hat{\boldsymbol{\theta}}_{\mathbf{m}},\, \varsigma_{j}\hat{\boldsymbol{\Sigma}}_{\mathbf{m}}),
\end{equation}
where $\sum_{j=1}^{N_{\mathrm{mix}}} w_{j} = 1$ and $f_{\mathrm{mvn}}(\boldsymbol{\theta}_{\mathbf{m}}^{*};\; \hat{\boldsymbol{\theta}}_{\mathbf{m}},\, \varsigma_{j}\hat{\boldsymbol{\Sigma}}_{\mathbf{m}})$ is a multivariate Gaussian density with mean vector $\hat{\boldsymbol{\theta}}_{\mathbf{m}}$ and covariance matrix $\varsigma_{j}\hat{\boldsymbol{\Sigma}}_{\mathbf{m}}$. The number of components $N_{\mathrm{mix}}$, the vector of weights $\mathbf{w} = (w_{1} \ddd w_{N_{\mathrm{mix}}})$, and the vector of scales $\boldsymbol{\varsigma} = (\varsigma_{1} \ddd \varsigma_{N_{\mathrm{mix}}})$ are tuning parameters that can be adjusted to help the sampling algorithm efficiently explore the posterior distribution.

The mean vector $\hat{\boldsymbol{\theta}}_{\mathbf{m}}$ and the base covariance matrix $\hat{\boldsymbol{\Sigma}}_{\mathbf{m}}$ must be set carefully to ensure that the proposal density for $\boldsymbol{\theta}_{\mathbf{m}}^{*}$ in \eqref{eq:sub_proposal} is a good approximation of the posterior density of the submodel
\begin{equation}
\label{eq:sub_posterior}
p(\boldsymbol{\theta}_{\mathbf{m}} \vbar \mathbf{m}^{*},\, \mathbf{r}) \propto p(\mathbf{r} \vbar \boldsymbol{\theta}_{\mathbf{m}},\, \mathbf{m}^{*})p(\boldsymbol{\theta}_{\mathbf{m}} \vbar \mathbf{m}^{*}).
\end{equation}
We first generate a sample path from a pilot MCMC run, targeting the submodel posterior in \eqref{eq:sub_posterior}, which is easily obtained by imposing zero restrictions on the conditional posterior of the full model $p(\boldsymbol{\theta} \vbar \mathbf{m}^{*},\, \mathbf{r})$. For this pilot run, we employ the \emph{random walk Metropolis} (RMW) kernel and a multivariate Gaussian proposal distribution, whose covariance matrix is adaptively estimated, up to a scaling factor, using the realised states of the chain, while the scaling factor is adaptively tuned to target an acceptance rate of 0.234, as in \citet{RobertsRosenthal2001}. After discarding an initial burn-in sample, we set $\hat{\boldsymbol{\theta}}_{\mathbf{m}}$ and $\hat{\boldsymbol{\Sigma}}_{\mathbf{m}}$ to be the sample mean and sample covariance matrix of the pilot chain. It is still acceptable if the stationary distribution of this pilot chain is not exactly \eqref{eq:sub_posterior} due to adaptation \citep{RobertsRosenthal2007}, since the sole purpose of the pilot chain is to learn good parameter values of $\hat{\boldsymbol{\theta}}_{\mathbf{m}}$ and $\hat{\boldsymbol{\Sigma}}_{\mathbf{m}}$. To reduce computational cost, we save the values of $\hat{\boldsymbol{\theta}}_{\mathbf{m}}$ and $\hat{\boldsymbol{\Sigma}}_{\mathbf{m}}$ for each newly visited knot configuration so that the rerunning of the RWM sampler is not required when the same configuration is revisited.

Note that conditional on $\mathbf{m}^{*}$, the proposal distribution is fixed, and the sampling algorithm effectively becomes an \emph{independence sampler} \citep{Tierney1994} targeting \eqref{eq:sub_posterior}. Similar to the method of importance sampling, the independence sampler is known to be efficient if the proposal density is a good approximation of the target while being heavier-tailed \citep{Roberts1996}. We set $N_{\mathrm{mix}} = 3$, $\mathbf{w} = (0.85,\, 0.1,\, 0.05)$, and $\boldsymbol{\varsigma} = (1,\, 10,\, 100)$, with the hope that Gaussian mixture proposal density is heavier-tailed than the target posterior in \eqref{eq:sub_posterior}, while we approximate the location, scale, and orientation of target by choosing $\hat{\boldsymbol{\theta}}_{\mathbf{m}}$ and $\hat{\boldsymbol{\Sigma}}_{\mathbf{m}}$ using the pilot chain.

\begin{figure}[h]
\centering
\begin{tikzpicture}
\tikzstyle{every node}=[circle,thick,draw,minimum size=1.15cm]
\tikzstyle{every path}=[thick,draw,->,>=latex]
\node (A) at (0,1) {$\mathbf{m}^{[i]}$};
\node (B) at (2,2) {$\boldsymbol{\theta}^{[i]}$};
\node (C) at (2,0) {$\mathbf{m}^{*}$};
\node (D) at (4,0) {$\boldsymbol{\theta}^{*}$};
\path (A) edge (B);
\path (A) edge (C);
\path (C) edge (D);
\end{tikzpicture}
\caption{A DAG encoding the set of conditional independencies implied by the proposal generating process described in step 2.}
\label{fig:dag_proposal}
\end{figure}
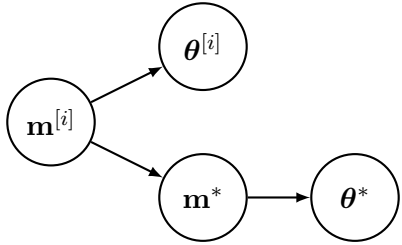

In step 3, as \eqref{eq:joint_proposal} is a density with respect to the natural measure on the product space $\mathbb{R}^{6+K} \times \{0,\,1\}^{K}$, the standard Metropolis-Hastings acceptance probability is a valid option for $u$ \citep{GruetRobert1997, Godsill2001, DellaportasForsterNtzoufras2002}, and is given by
\begin{equation}
\label{eq:acc_prob_1}
\begin{aligned}
u &= \min \left[\frac{p(\boldsymbol{\theta}^{*},\, \mathbf{m}^{*} \vbar \mathbf{r}) q(\boldsymbol{\theta}^{[i]},\, \mathbf{m}^{[i]} \vbar \boldsymbol{\theta}^{*},\, \mathbf{m}^{*})}{p(\boldsymbol{\theta}^{[i]},\, \mathbf{m}^{[i]} \vbar \mathbf{r}) q(\boldsymbol{\theta}^{*},\, \mathbf{m}^{*} \vbar \boldsymbol{\theta}^{[i]},\, \mathbf{m}^{[i]})},\, 1 \right] \\
&= \min \left[\frac{p(\boldsymbol{\theta}^{*},\, \mathbf{m}^{*} \vbar \mathbf{r}) q(\boldsymbol{\theta}^{[i]} \vbar \mathbf{m}^{[i]},\, \boldsymbol{\theta}^{*},\, \mathbf{m}^{*}) q(\mathbf{m}^{[i]} \vbar \boldsymbol{\theta}^{*},\, \mathbf{m}^{*})}{p(\boldsymbol{\theta}^{[i]},\, \mathbf{m}^{[i]} \vbar \mathbf{r}) q(\boldsymbol{\theta}^{*} \vbar \mathbf{m}^{*},\, \boldsymbol{\theta}^{[i]},\, \mathbf{m}^{[i]}) q(\mathbf{m}^{*} \vbar \boldsymbol{\theta}^{[i]},\, \mathbf{m}^{[i]})},\, 1 \right].
\end{aligned}
\end{equation}
The set of conditional independencies implied by the construction of the proposal distribution in step 2 is represented graphically in Figure~\ref{fig:dag_proposal}, using a directed acyclic graph (DAG). Using the DAG, the expression for $u$ can be simplified to
\begin{equation}
\label{eq:acc_prob_2}
u = \min \left[ \frac{p(\boldsymbol{\theta}^{*},\, \mathbf{m}^{*} \vbar \mathbf{r}) q(\boldsymbol{\theta}^{[i]} \vbar \mathbf{m}^{[i]}) q(\mathbf{m}^{[i]} \vbar \mathbf{m}^{*})}{p(\boldsymbol{\theta}^{[i]},\, \mathbf{m}^{[i]} \vbar \mathbf{r}) q(\boldsymbol{\theta}^{*} \vbar \mathbf{m}^{*}) q(\mathbf{m}^{*} \vbar \mathbf{m}^{[i]})},\, 1 \right].
\end{equation}
Using the fact that the proposal density for the knot configuration $\mathbf{m}^{*}$, given by \eqref{eq:m_proposal}, is symmetric, the acceptance probability is further reduced to
\begin{equation}
\label{eq:acc_prob_3}
u = \min \left[ \frac{p(\boldsymbol{\theta}^{*},\, \mathbf{m}^{*} \vbar \mathbf{r}) q(\boldsymbol{\theta}^{[i]} \vbar \mathbf{m}^{[i]})}{p(\boldsymbol{\theta}^{[i]},\, \mathbf{m}^{[i]} \vbar \mathbf{r}) q(\boldsymbol{\theta}^{*} \vbar \mathbf{m}^{*})},\, 1 \right].
\end{equation}

\FloatBarrier

\section{Simulation study}
\label{sec:simulation}
A simulation study is conducted to investigate the comparative performance of the proposed SP-GARCH model and the popular parametric alternatives. The parametric models considered are the first order GARCH, GJR-GARCH (GJR), and Beta-t-GARCH (Beta-t). The aim is to investigate whether the SP-GARCH model is able to offer more accurate estimates of the conditional volatilities $\{\sigma_{t}\colon\, t \in \{1 \ddd T\}\}$ such that $\sigma_{t} = [\Var(r_{t} \vbar \mathcal{F}_{t-1})]^{1/2}$ for certain family of processes followed by $\{r_{t}\}$. The focus is on comparing structural features of the estimated coefficient functions and measures of loss in both the in-sample and out-of-sample estimates of conditional volatilities. %Additionally, the properties of the estimates of the parameters $\nu$, $\mu$, and $\omega$ are examined in terms of bias and variability. These parameters are common to all the models under comparison.

The model definitions of the comparing parametric models are given by \eqref{eq:spgarch} and Table~\ref{tab:gfunc_para}. The innovation distribution for each model is $F_{\mathrm{sdt},\nu}$. The parameters are constrained to ensure the stationarity of $\{r_{t}\}$ and the positivity of $\{\sigma_{t}^{2}\}$, by truncating the likelihood, similar to \eqref{eq:likelihood}. For each parametric model, the flat prior is used for most of the parameters with an exception of $\nu$, where the prior is proportional to $\nu^{-2}$.

Independent sequences of $T$ observations are generated from four data generating processes (DGPs), labelled DGP~1, DGP~2, DGP~3, and DGP~4. The DGPs can be written in the functional coefficient form of \eqref{eq:spgarch}, with DGP~$i$, for $i \in \{1 \ddd 4\}$, given by
\begin{equation}
\label{eq:dgp}
\begin{aligned}
r_{t} &= \sigma_{t} \epsilon_{i,t}, \\
\sigma^{2}_{t} &= 0.1 + g_{i}(\epsilon_{i,t-1}) \sigma^{2}_{t-1},
\end{aligned}
\end{equation}
where $\{\epsilon_{i,t}\}$ is an i.i.d.~sequence from $F_{\mathrm{sdt},8}$ for $i \in \{1,2\}$, and from $F_{\mathrm{sdt},5}$ for $i \in \{3,4\}$. The coefficient functions $g_{1}(\cdot) \ddd g_{4}(\cdot)$ are specified as follows.
\begin{equation}
\label{eq:g1to4}
\begin{aligned}
g_{1}(\epsilon) &= 1.1 - 0.48(\epsilon + 0.77)^{2}_{+} + 0.58(\epsilon + 0.473)^{2}_{+}; \\
g_{2}(\epsilon) &= 0.85 + 0.1\epsilon^{2}; \\
g_{3}(\epsilon) &= 0.82 + 0.15[6\epsilon^{2}/(3 + \epsilon^{2})]; \\
g_{4}(\epsilon) &= 0.8 + [0.1 + 0.15 I_{(-\infty,0)}(\epsilon)]\epsilon^{2}.
\end{aligned}
\end{equation}
The function $g_{1}(\cdot)$ is a quadratic spline with two knots at -0.77 and -0.473. The motivation for $g_{1}(\cdot)$ is to construct a coefficient function with structural features that cannot be fully captured by most parametric GARCH models, yet are not completely unrealistic from an empirical perspective. For example, $g_{1}(\cdot)$ implies that a relatively small negative return impacts the next day conditional volatility differently from a small positive return of the same magnitude, while the impact of a negative return stops to grow with its magnitude once the return is beyond a certain threshold. The functions $g_{2}(\cdot)$, $g_{3}(\cdot)$, and $g_{4}(\cdot)$ correspond to the coefficient functions of the first order GARCH, Beta-t, and GJR models, respectively. It can be checked that all DGPs satisfy the covariance stationarity condition in \eqref{eq:stationarity}, with $\E[g_{1}(\epsilon_{1,t})] = 0.977$, $\E[g_{2}(\epsilon_{2,t})] = 0.95$, $\E[g_{3}(\epsilon_{3,t})] = 0.97$, and $\E[g_{4}(\epsilon_{4,t})] = 0.975$.

We generate $N_{\mathrm{sim}} = 500$ independent sequences of length $T = 4001$ from each of the two DGPs. The posterior mean estimates of the conditional volatilities (as well as other quantities of interest presented in this section) are computed using the first 4000 observations of each generated sequence; the last observation of each sequence is left out for quantifying the out-of-sample performance.

To compute the posterior mean estimates for the SP-GARCH model via \eqref{eq:phi_mcmc}, the adaptive MCMC sampling algorithm detailed in Section~\ref{sec:mcmc} is employed. The sampling algorithm is run for $5.5 \times 10^{5}$ iterations, where the first $0.5 \times 10^{5}$ burn-in iterations are ignored when computing the posterior mean estimates. The potential knots $k_{1} \ddd k_{9}$ are chosen as quantiles of the standardised Student-t distribution with 8 degrees-of-freedom, at levels $0.1 \ddd 0.9$, respectively; the step-size parameter $\pi$ in \eqref{eq:m_proposal} is set to 0.1; the shrinkage parameter $\sigma^{2}_{\beta}$ in \eqref{eq:prior_theta} is set to 2500.

To compute the posterior mean estimates for the parametric models, we use an independence sampler with a multivariate Gaussian mixture proposal configured in the same way as \eqref{eq:sub_proposal}. The mean vector and the base covariance matrix of the Gaussian mixture are set to the sample mean and covariance matrix of a pilot chain generated by the same adaptive RWM sampler as the one used in step 2(b) of the sampling algorithm in Section~\ref{sec:mcmc}. The independence sampler is shown to work well for parametric GARCH models in \citet{ChenSo2006} and \citet{ChenGerlachLu2012}.

\begin{figure}[h]
\hspace{-2.4cm}
\includegraphics[width = 1.3 \textwidth]{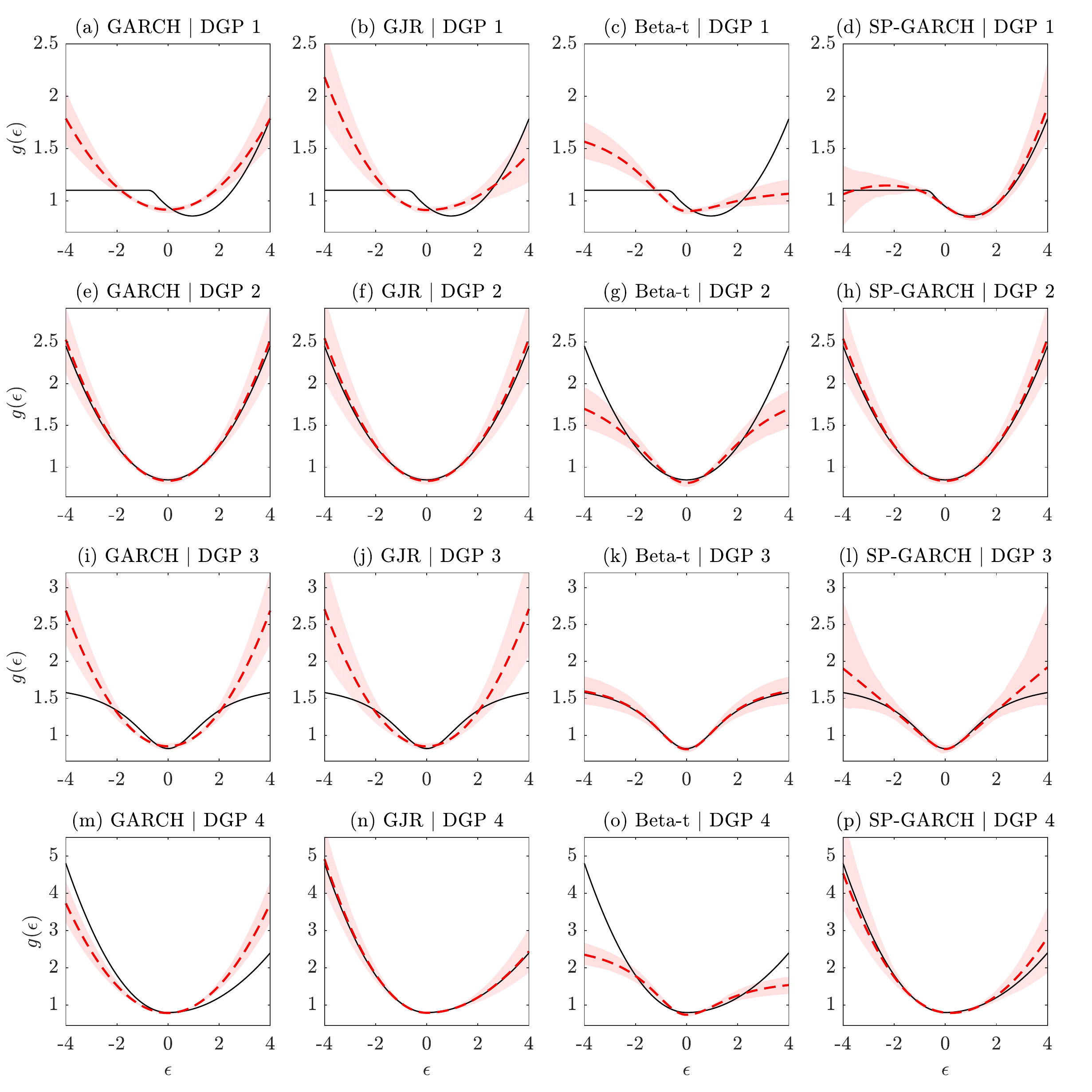}
\caption{Each plot contains the MC mean of the 500 posterior mean estimates of the coefficient functions (red dashed line), the pointwise 95\% MC confidence band (pink shaded area), and the true function (black line).}
\label{fig:gfunc_sim}
\end{figure}

The Monte Carlo (MC) mean of the 500 posterior mean estimates of the coefficient functions is plotted for each model in Figure~\ref{fig:gfunc_sim}, along with the 95\% MC confidence band. Both the MC mean and the MC confidence band are computed pointwise using an equally spaced grid of points over $[-4,\, 4]$ with a spacing of 0.01. For DGP 1, only the SP-GARCH is able to capture the essential features of the true coefficient function. It is interesting to see that both the GJR and Beta-t models are trying to capture the asymmetry of $g_{1}(\cdot)$ near zero, where most of the innovations are found, while missing the important features for large innovations. For DGP 2, which corresponds to the the GARCH model, all the models except for the Beta-t are able to approximate the simple quadratic function. The Beta-t model is performing poorly due to the fact that its coefficient function deviates from the quadratic from if the innovation distribution is heavier tailed than the Gaussian. Conversely, when the true model is Beta-t, as for DGP 3, the quadratic models (GARCH and GJR) provide poor approximations as $g_{3}(\cdot)$ increases faster than quadratic for small innovations, but grows much slower than quadratic for large innovations. The SP-GARCH is able to offer a good approximation to $g_{3}(\cdot)$ except for the tails of the innovation distribution where the data is sparse. Lastly, when the true model is GJR, SP-GARCH is able to learn the shape of $g_{4}(\cdot)$ reasonably well. Expectedly, the approximation becomes increasingly biased as $\epsilon$ moves deeper into the tails of the innovation distribution.

\FloatBarrier

To compare the quality of the posterior mean estimates of the conditional volatilities between the models for each of the two DGPs, both the in-sample and out-of-sample versions of the $L_{p}$ loss are considered. For the $i$th simulated sequence, the in-sample $L_{p}$ loss is defined, for $p \ge 1$, as
\begin{equation}
\label{eq:lp_in}
L_{p,i}^{(\mathrm{In})} = \left[\frac{1}{T-1}\sum_{t=1}^{T-1}|\hat{\sigma}_{i,t} - \sigma_{i,t}|^{p}\right]^{1/p},
\end{equation}
where $\hat{\sigma}_{i,t}$ is the posterior mean estimate of the volatility for $t$th observation of the $i$th simulated sequence, and $\sigma_{i,t}$ is the corresponding true volatility; for the holdout observations across different sequences, the out-of-sample $L_{p}$ loss is defined, for $p \ge 1$, as
\begin{equation}
\label{eq:lp_out}
L_{p}^{(\mathrm{Out})} = \left[\frac{1}{N_{\mathrm{sim}}}\sum_{i=1}^{N_{\mathrm{sim}}}|\hat{\sigma}_{i,T} - \sigma_{i,T}|^{p}\right]^{1/p}.
\end{equation}
Notice that the out-of-sample $L_{p}$ loss is calculated cross-sectionally rather than time-series-wise. The cases $p \in \{1, 2\}$ are considered in this study: When $p = 2$, the loss measure $L_{p}^{(\cdot)}$ corresponds to the usual root-mean-square-error (RMSE); when $p = 1$, $L_{p}^{(\cdot)}$ corresponds to the usual mean-absolute-error (MAE).

The MC mean of the 500 realisations of the in-sample $L_{p}$ loss is reported for each $p \in \{1, 2\}$, each model, and each DGP in Table~\ref{tab:loss_sim_in}, where $\bar{L}_{p}^{(\mathrm{In})}$ denotes $(1/500)\sum_{i=1}^{500}L_{p,i}^{(\mathrm{In})}$. Table~\ref{tab:loss_sim_out} is the out-of-sample analogue of Table~\ref{tab:loss_sim_in}. We make some observations: (i) Comparing Table~\ref{tab:loss_sim_in} with Figure~\ref{fig:gfunc_sim}, it is evident that conditional volatility estimates are sensitive to mis-specification of the coefficient function. (ii) Compared to the parametric models, SP-GARCH is robust to such mis-specification as it can adapt to a wide range of coefficient functions. (iii) For certain class of coefficient functions, the incurred efficiency loss (by not using the correct parametric model) is very small. For example, when the true DGP is GARCH, SP-GARCH is as efficient as the GJR model, given the fact that GJR has only one more degree-of-freedom than that of GARCH, and that SP-GARCH is a much more flexible model. (iv) The relative accuracy of SP-GARCH is retained out-of-sample. This suggests that the accuracy gain is not caused by overfitting.

\begin{table}[h]
\centering
\begin{tabular}{lcccccccc}
\toprule
 & \multicolumn{2}{c}{DGP 1} & \multicolumn{2}{c}{DGP 2} & \multicolumn{2}{c}{DGP 3} & \multicolumn{2}{c}{DGP 4} \\ 
\cmidrule(lr){2-3} \cmidrule(lr){4-5} \cmidrule(lr){6-7} \cmidrule(lr){8-9}
 & $\bar{L}_{2}^{(\mathrm{In})}$ & $\bar{L}_{1}^{(\mathrm{In})}$ & $\bar{L}_{2}^{(\mathrm{In})}$ & $\bar{L}_{1}^{(\mathrm{In})}$ & $\bar{L}_{2}^{(\mathrm{In})}$ & $\bar{L}_{1}^{(\mathrm{In})}$ & $\bar{L}_{2}^{(\mathrm{In})}$ & $\bar{L}_{1}^{(\mathrm{In})}$ \\ 
\midrule
GARCH & 0.296 & 0.229 & 0.035 & 0.026 & 0.228 & 0.124 & 0.196 & 0.115 \\ 
GJR & 0.292 & 0.219 & 0.041 & 0.030 & 0.229 & 0.126 & 0.063 & 0.039 \\ 
Beta-t & 0.267 & 0.201 & 0.087 & 0.054 & 0.058 & 0.044 & 0.268 & 0.116 \\ 
SP-GARCH & 0.093 & 0.069 & 0.040 & 0.030 & 0.102 & 0.066 & 0.096 & 0.055 \\ 
\bottomrule
\end{tabular}
\caption{MC means of the 500 in-sample $L_{p}$ loss values between estimated and true conditional volatilities.}
\label{tab:loss_sim_in}
\end{table}
\FloatBarrier

\begin{table}[h]
\centering
\begin{tabular}{lcccccccc}
\toprule
 & \multicolumn{2}{c}{DGP 1} & \multicolumn{2}{c}{DGP 2} & \multicolumn{2}{c}{DGP 3} & \multicolumn{2}{c}{DGP 4} \\ 
\cmidrule(lr){2-3} \cmidrule(lr){4-5} \cmidrule(lr){6-7} \cmidrule(lr){8-9}
 & $L_{2}^{(\mathrm{Out})}$ & $L_{1}^{(\mathrm{Out})}$ & $L_{2}^{(\mathrm{Out})}$ & $L_{1}^{(\mathrm{Out})}$ & $L_{2}^{(\mathrm{Out})}$ & $L_{1}^{(\mathrm{Out})}$ & $L_{2}^{(\mathrm{Out})}$ & $L_{1}^{(\mathrm{Out})}$ \\ 
\midrule
GARCH & 0.308 & 0.233 & 0.037 & 0.026 & 0.265 & 0.138 & 0.171 & 0.107 \\ 
GJR & 0.319 & 0.227 & 0.042 & 0.030 & 0.266 & 0.137 & 0.073 & 0.042 \\ 
Beta-t & 0.266 & 0.205 & 0.074 & 0.050 & 0.062 & 0.043 & 0.204 & 0.106 \\ 
SP-GARCH & 0.094 & 0.068 & 0.041 & 0.029 & 0.137 & 0.069 & 0.096 & 0.056 \\ 
\bottomrule
\end{tabular}
\caption{Out-of-sample $L_{p}$ loss values between estimated and true conditional volatilities.}
\label{tab:loss_sim_out}
\end{table}
\FloatBarrier

\section{Empirical study}
\label{sec:empirical}
As an empirical study, the SP-GARCH model is applied to the daily returns of five stock market indices and five individual stocks. The aim is to gain insights into the conditional volatility process of each series by analysing the estimated coefficient function. The five stock market indices are: S\&P 500 (US), FTSE 100 (UK), DAX (Germany), Nikkei 225 (Japan), and Hang Seng (Hong Kong); the five individual stocks are: Apple, ARM, Intel, Nvidia, and SanDisk. A sample of daily log-returns is obtained for each index or stock. Most of the samples start in January 1996, with exceptions of ARM and Nvidia, for which the samples start in April 1998 and January 1999, respectively. All of the samples end in November 2015. Nvidia has the smallest sample size of 4236 observations, while FTSE has the largest sample size of 5176 observations.

For each sample, the posterior mean estimate of the coefficient function and the 95\% credible band are computed using the MCMC sampling algorithm detailed in Section~\ref{sec:mcmc}. The sequence of potential knots, the configuration of the sampling algorithm, and the value of the hyperparameter are the same as those used for the simulation study in Section~\ref{sec:simulation}.

\begin{figure}[h]
\centering
\includegraphics[width = 0.85 \textwidth]{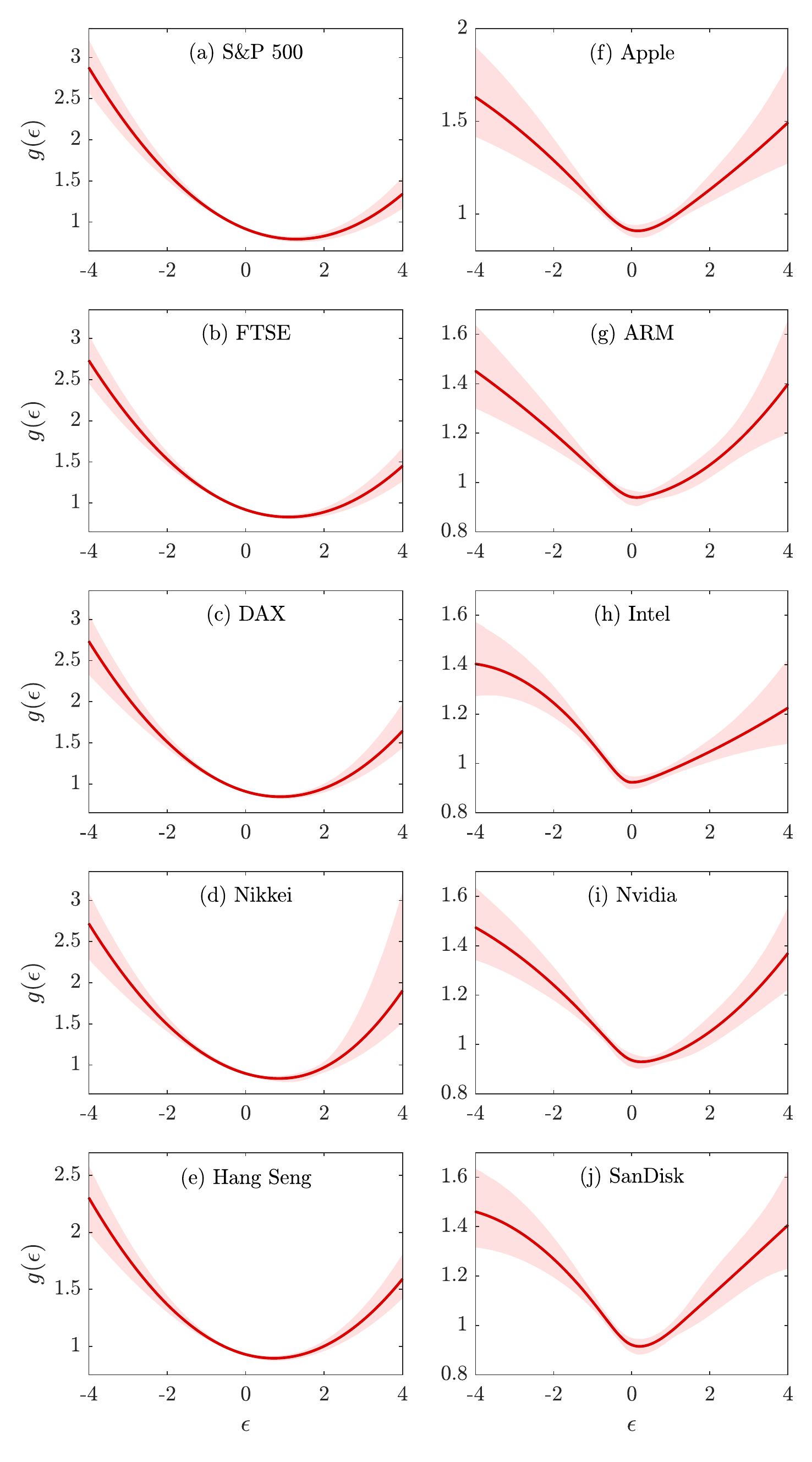}
\caption{Posterior mean estimates (red solid lines) and pointwise 95\% credible bands (pink shaded areas) of the coefficient functions for the stock indices and individual stocks.}
\label{fig:gfunc_emp}
\end{figure}

In Figure~\ref{fig:gfunc_emp}, the posterior mean estimate of the coefficient function is plotted for each of the ten samples, together with the pointwise 95\% credible band. It is striking to see that the samples can be easily separated into two groups based on the structural features of the coefficient functions; there appear to be a clear distinction between the coefficient functions for the stock indices and those for the individual stocks. For the stock indices, the functions appear to be quadratic, with asymmetries created by shifting the minimums away from zero; for the individual stocks, the functions appear to be piecewise-linear-like, with asymmetries created by having slopes with varying degrees of steepness for negative and positive innovations. There also appear to be a ``threshold effect" for large negative innovations for some individual stocks such as Intel and SanDisk.

\FloatBarrier

The posterior summaries of $\nu$ and $\mu$ are reported in Table~\ref{tab:est_emp}. For each parameter-sample combination, we report the posterior mean (Mean) and the 95\% credible interval (Lower, Upper). One can observe that the innovation distributions for the individual stocks are heavier tailed than those for the stock indices, as indicated by the estimates of $\nu$. The estimates of $\mu$ indicate that the long-run (unconditional) means of the daily return processes are all positive but small in magnitude, with Apple generating the largest expected daily return of around 0.1\%.

\begin{table}[h]
\centering
\begin{tabular}{lcccccc}
\toprule
& \multicolumn{3}{c}{$\nu$} & \multicolumn{3}{c}{$\mu$} \\
\cmidrule(lr){2-4} \cmidrule(lr){5-7}
& Mean & Lower & Upper & Mean & Lower & Upper \\ 
\midrule
S\&P 500 & 8.935 & 7.109 & 11.363 & 0.029 & 0.006 & 0.052 \\ 
FTSE & 11.189 & 8.546 & 14.958 & 0.010 & -0.012 & 0.032 \\ 
DAX & 11.498 & 8.688 & 15.576 & 0.052 & 0.022 & 0.081 \\ 
Nikkei & 11.138 & 8.504 & 14.782 & 0.013 & -0.021 & 0.046 \\ 
Hang Seng & 7.304 & 5.980 & 9.045 & 0.034 & 0.003 & 0.065 \\ 
Apple & 5.023 & 4.407 & 5.731 & 0.118 & 0.062 & 0.173 \\ 
ARM & 4.666 & 4.089 & 5.338 & 0.061 & -0.006 & 0.126 \\ 
Intel & 6.861 & 5.840 & 8.095 & 0.049 & 0.002 & 0.096 \\ 
Nvidia & 4.930 & 4.318 & 5.648 & 0.043 & -0.027 & 0.114 \\ 
SanDisk & 4.191 & 3.732 & 4.709 & 0.049 & -0.022 & 0.120 \\ 
\bottomrule
\end{tabular}
\caption{Posterior summaries of $\nu$ and $\mu$ for stock indices and individual stocks.}
\label{tab:est_emp}
\end{table}
\FloatBarrier

\begin{figure}[h]
\centering
\includegraphics[width = 0.7 \textwidth]{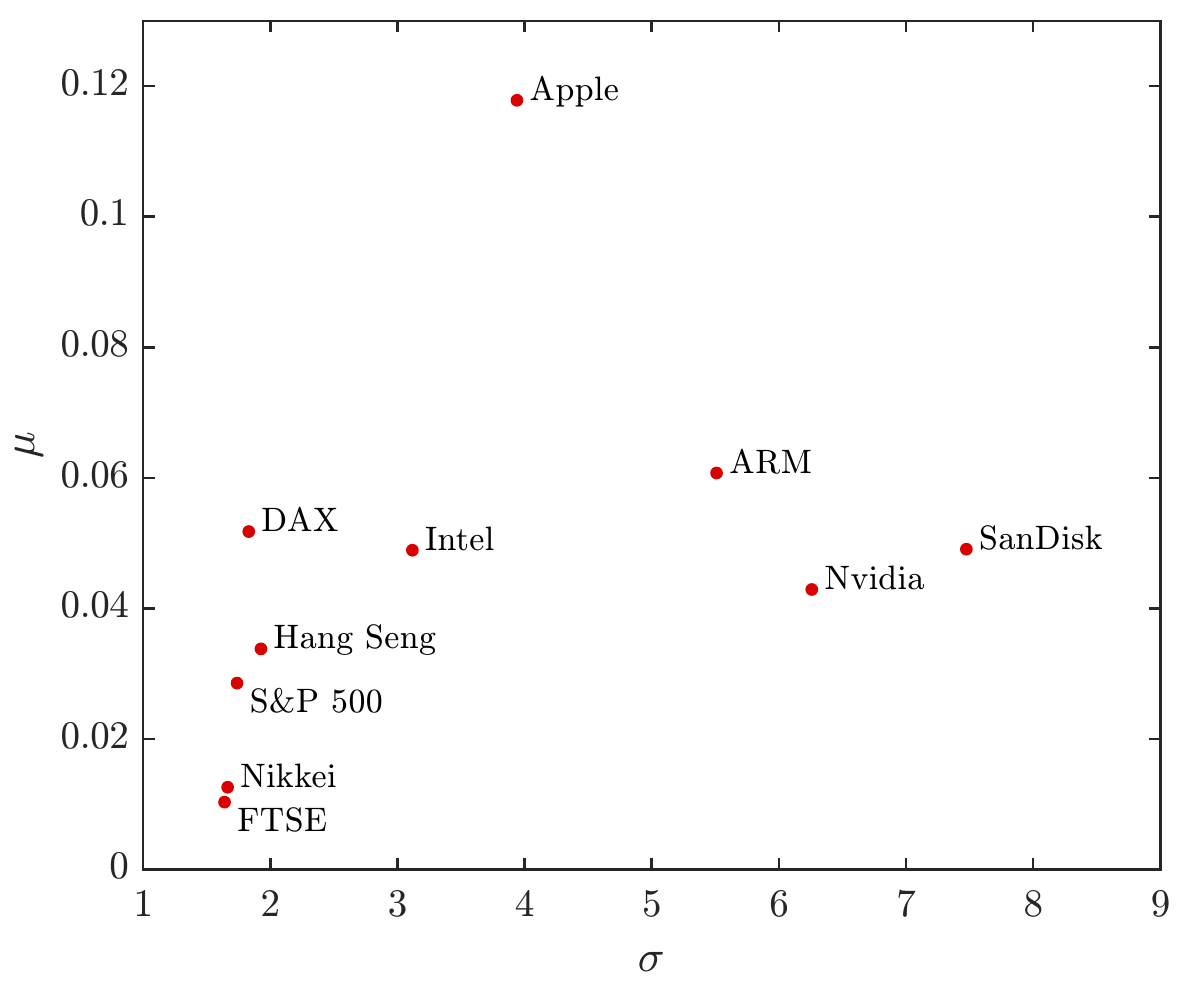}
\caption{Scatter plot showing the posterior means of the unconditional expected returns of the stock indices and individual stocks plotted against (the posterior means of) the corresponding unconditional volatilities.}
\label{fig:sd_vs_mn}
\end{figure}

One may be interested in estimating the long-run (unconditional) volatilities of the return processes as quantitative measures of the long-run daily risk; $\sigma = \Var(r_{t})^{1/2}$. This can be done using \eqref{eq:uncvar} and \eqref{eq:ex_g} for the SP-GARCH model. In Figure~\ref{fig:sd_vs_mn}, a scatter plot is generated by plotting each sample in the product space of unconditional volatility and unconditional mean; $\{\sigma\} \times \{\mu\}$. Such plot is useful for comparing the long-run daily risk-return profiles of the samples, which could correspond to assets in an investment portfolio. It can be seen from the plot that both DAX and Apple show favourable risk-return characteristics compared to the rest of the samples.

\FloatBarrier

\begin{figure}[h]
\centering
\includegraphics[width = 1.0 \textwidth]{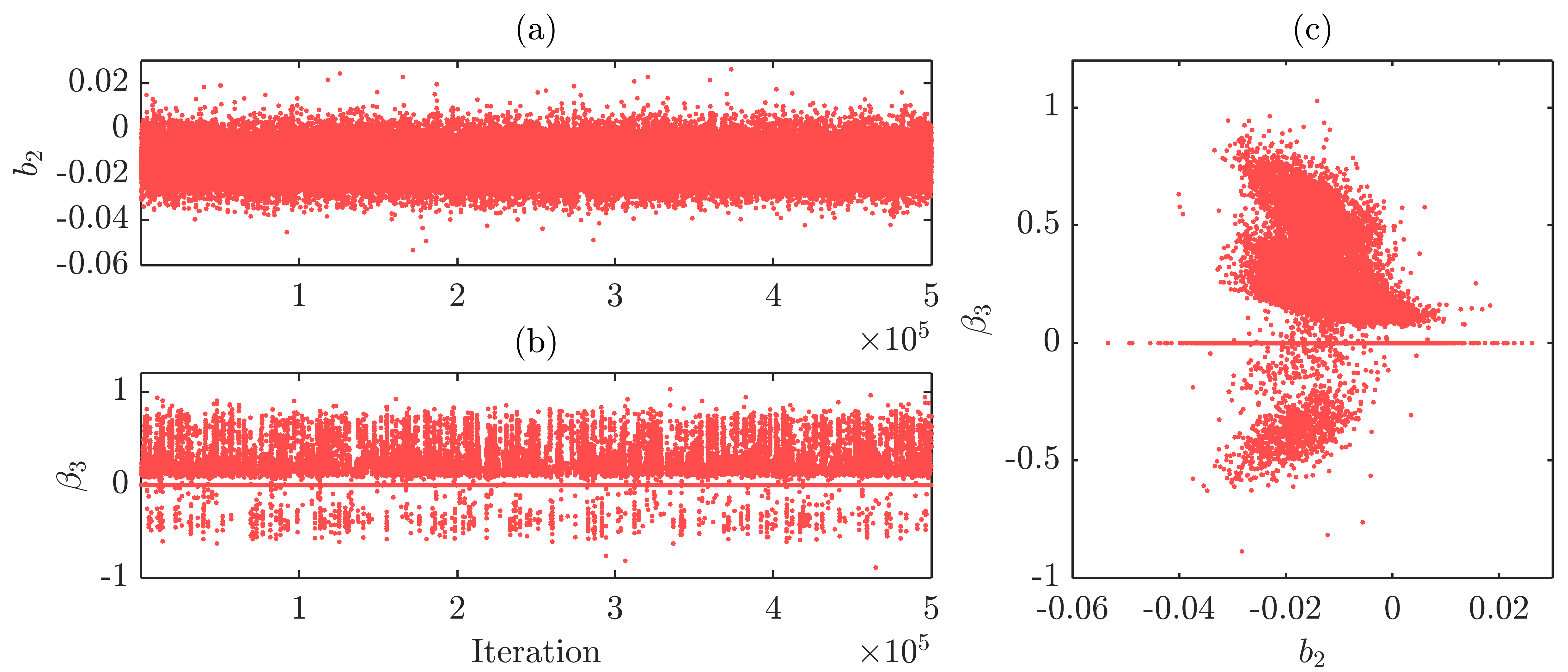}
\caption{Trace plots and scatter plots of the Markov chain showing realisations from the marginal posterior distributions and the joint marginal posterior distribution of $b_{2}$ and $\beta_{3}$ for Apple.}
\label{fig:chain_theta}
\end{figure}

As an illustration of the mixing properties of the constructed Markov chain when the sampling algorithm is applied to real data, the realised values of the chain in dimensions $b_{2}$ and $\beta_{3}$, after discarding the initial burn-in iterations, are plotted against the iteration counts minus $5.5 \times 10^{5}$ in panels (a) and (b) of Figure~\ref{fig:chain_theta}, when the posterior distribution is conditional on the Apple data. It appears that the chain is settled down to rapid mixing in both dimensions. In panel (b), it is reassuring to see that the algorithm appears to be able to effectively switch between multiple modes including a point mass at exactly zero. The scatter plot in panel (c) reveals the complex landscape of the joint marginal posterior distribution of $b_{2}$ and $\beta_{3}$.

\FloatBarrier

\begin{figure}[h]
\centering
\includegraphics[width = 1.0 \textwidth]{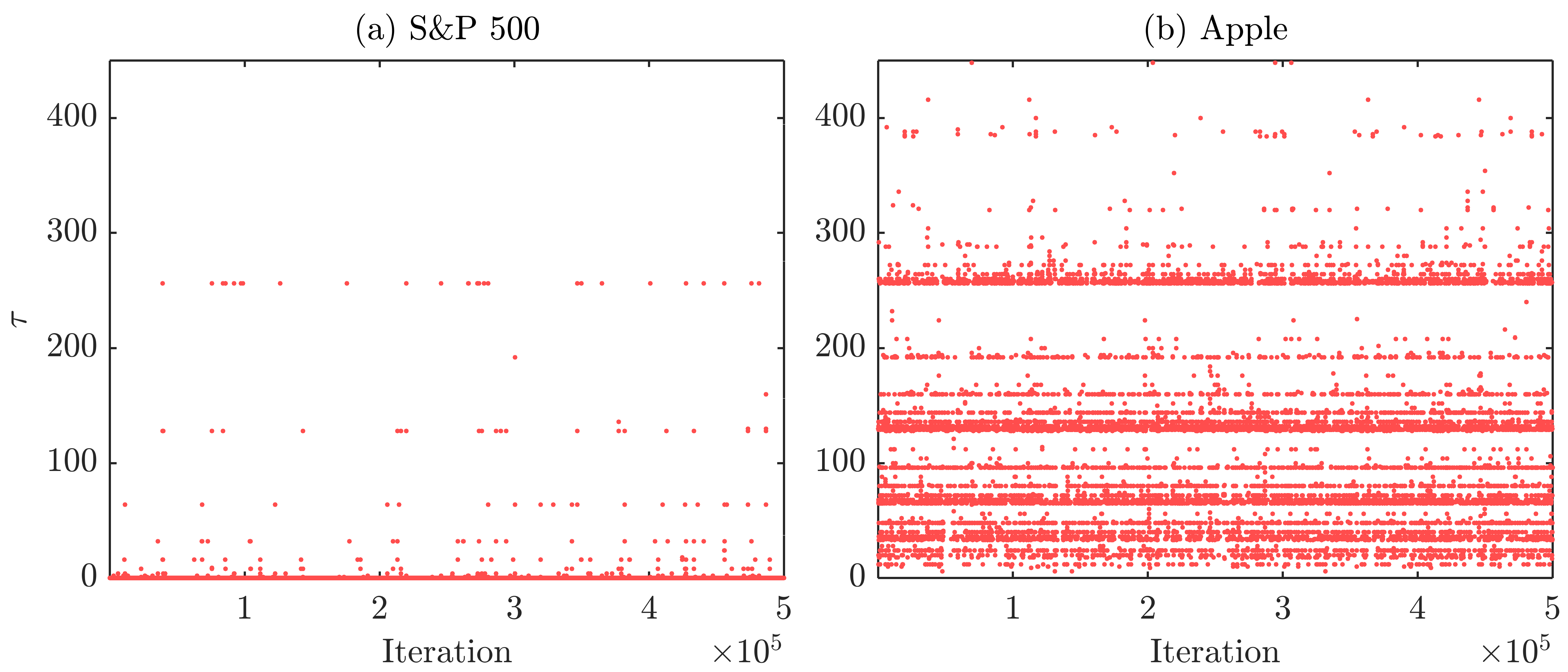}
\caption{Trace plots of the Markov chain in model space showing samples from the marginal posterior distribution of $\tau = n_{0}(\mathbf{m})$ for S\&P 500 and Apple, where $n_{0}\colon \{0,\, 1\}^{K} \rightarrow \mathbb{N}_{0}$ converts a binary vector to a nonnegative decimal integer.}
\label{fig:chain_model}
\end{figure}

It is useful to also examine the mixing properties of the Markov chain of the binary vectors, i.e., the Markov chain in model space. One way to visualise the sample path (or trace) of such chain is to first map each realisation of the $K$-dimensional binary vector $\mathbf{m}^{[i]}$ to a scalar $\tau^{[i]}$, for $i \in \{1 \ddd N\}$, such that the mapping is one-to-one, and then plot $\tau^{[i]}$ against $i$, for $i \in \{1 \ddd N\}$. One such mapping is easily found by treating each binary vector as a $K$-bit binary number. The mapped scalar is then the decimal equivalent of the binary number. To illustrate the idea, this mapping is used to generate the trace plots in Figure~\ref{fig:chain_model} for S\&P 500 and Apple. For both samples, the chain appears to be rapidly mixing in model space. 

\FloatBarrier

A simple measure of model complexity for the SP-GARCH model is the number of knots chosen for the regression spline, given by
\begin{equation}
\label{eq:km}
K_{\mathbf{m}} = \sum_{i=1}^{K}m_{i}.
\end{equation}
Using the realised binary vectors of the Markov chain, one can easily compute the marginal posterior probability of $K_{\mathbf{m}}$. Such posterior probabilities are reported in Table~\ref{tab:post_km} for each sample for $K_{\mathbf{m}} \le 3$. For stock indices, the most favoured model is the one without any knots; for individual stocks, more complex models are preferred, as most of the posterior mass is on $K_{\mathbf{m}} = 2$.

\begin{table}[h]
\centering
\begin{tabular}{lcccc}
    \toprule
    $K_{\mathbf{m}}$ & 0     & 1     & 2     & 3 \\
    \midrule
    S\&P 500 & 0.991 & 0.009 & 0.000 & 0.000 \\
    FTSE  & 0.987 & 0.013 & 0.000 & 0.000 \\
    DAX   & 0.935 & 0.064 & 0.001 & 0.000 \\
    Nikkei & 0.762 & 0.236 & 0.002 & 0.000 \\
    Hang Seng & 0.957 & 0.042 & 0.001 & 0.000 \\
    Apple & 0.000 & 0.009 & 0.976 & 0.015 \\
    ARM   & 0.000 & 0.313 & 0.672 & 0.015 \\
    Intel & 0.000 & 0.004 & 0.982 & 0.014 \\
    Nvidia & 0.000 & 0.135 & 0.851 & 0.014 \\
    SanDisk & 0.000 & 0.018 & 0.965 & 0.017 \\
    \bottomrule
\end{tabular}
\caption{Marginal posterior probabilities of model complexity measured by the number of nonzero knot coefficients.}
\label{tab:post_km}
\end{table}
\FloatBarrier

It is possible that the structural features of the estimated coefficient functions shown in Figure~\ref{fig:gfunc_emp} are not unique, due to multimodality of the joint posterior distributions, and that our MCMC sampler is not effectively exploring all the modes. This possibility can potentially render our conclusions concerning the shapes of the coefficient functions invalid. For example, there could be a parametric GARCH model that fits the data equally well as the SP-GARCH, but giving a structurally different estimate of the coefficient function. One way to guard against this possibility is to compare SP-GARCH against the parametric models using a measure of model adequacy. We need to ensure that SP-GARCH is indeed a more adequate description of the data when the inference about the coefficient function is different under SP-GARCH.

We perform a model comparison study based on the deviance information criterion (DIC) of \citet{SpiegelhalterEtAl2002} which can be considered as a Bayesian generalisation of the Akaike information criterion (AIC) \citep{Akaike1973}. Successful applications of the DIC include \citet{BergMeyerYu2004} and \citet{ChanGrant2016} for comparing stochastic volatility models, and \citet{Ardia2009} for comparing Markov-switching GARCH models. When computing the DIC, one obtains both a measure of model fit and model complexity. The model with the smallest DIC achieves the best fit-complexity trade-off.

Usually, a flexible semiparametric model like the SP-GARCH would not perform well under the DIC due to its high complexity. However, the proposed BMA approach will cause the posterior mass to concentrate over simpler models when more complex models do not significantly improve the likelihood (i.e., model fit). It would thus be interesting to investigate how SP-GARCH performs against the parametric models when both fit and complexity are taken into consideration.

A few studies have investigated the definition and the computation of the DIC for latent variable models \citep{CeleuxEtAl2006, LiZengYu2013, Ardia2009, ChanGrant2016}, however the DIC has not been examined in the context of BMA. For example, it is not clear how one may compute the DIC using the output of the reversible jump MCMC sampler of \citet{Green1995}, where the models are sampled together with their parameters. A natural generalisation of the DIC in the context of BMA may be the \emph{averaged} DIC defined as follows (using the more general notation in Section~\ref{sec:bma}).
\begin{equation}
\label{eq:dic_ave}
\mathrm{DIC}_{\mathrm{ave}} = \E_{\tau}(\mathrm{DIC}_{\tau} \vbar \mathbf{r}) = \sum_{\tau=1}^{M} \mathrm{DIC}_{\tau} p(\tau \vbar \mathbf{r}),
\end{equation}
where $\tau \in \{1 \ddd M\}$ is the model index, $p(\tau \vbar \mathbf{r})$ is the posterior probability of the model $\tau$. The DIC for model $\tau$, denoted by $\mathrm{DIC}_{\tau}$, is the usual single model DIC given by
\begin{equation}
\label{eq:dic}
\mathrm{DIC}_{\tau} = \bar{D}_{\tau} + p_{D}^{\tau}.
\end{equation}
The first term is the conditional posterior mean of the deviance
\begin{equation}
\label{eq:dbar_tau}
\bar{D}_{\tau} = \E_{\boldsymbol{\theta}_{\tau}}[D(\boldsymbol{\theta}_{\tau},\, \tau) \vbar \tau,\, \mathbf{r}],
\end{equation}
which can be considered as a Bayesian measure of model fit. The deviance is defined as
\begin{equation}
\label{eq:d}
D(\boldsymbol{\theta}_{\tau},\, \tau) = -2 \log p(\mathbf{r} \vbar \boldsymbol{\theta}_{\tau},\, \tau),
\end{equation}
where $p(\mathbf{r} \vbar \boldsymbol{\theta}_{\tau},\, \tau)$ is the likelihood. The second term in \eqref{eq:dic} is the effective number of parameters -- a measure of model complexity, and is defined as
\begin{equation}
\label{eq:pd_tau}
p_{D}^{\tau} = \bar{D}_{\tau} - D(\bar{\boldsymbol{\theta}}_{\tau},\, \tau),
\end{equation}
where $\bar{\boldsymbol{\theta}}_{\tau}$ is the conditional posterior mean of $\boldsymbol{\theta}_{\tau}$; $\bar{\boldsymbol{\theta}}_{\tau} = \E_{\boldsymbol{\theta}_{\tau}}(\boldsymbol{\theta}_{\tau} \vbar \tau,\, \mathbf{r})$. It can be seen from the definition in \eqref{eq:dic_ave} that, for any consistent model selection procedure, the averaged DIC in \eqref{eq:dic_ave} is asymptotically equivalent to the single model DIC of \citet{SpiegelhalterEtAl2002}.

It is straightforward to compute an estimate of $\mathrm{DIC}_{\mathrm{ave}}$ whenever the posterior sample $\{(\boldsymbol{\theta}_{\tau}^{[i]},\, \tau^{[i]})\colon\, i \in \{1 \ddd N\}\}$ is available. Let the index set $\mathcal{I}_{\tau} \subseteq \{1 \ddd N\}$ be defined as follows.
\begin{equation}
\label{eq:set_i_tau}
\mathcal{I}_{\tau} = \left\{i_{1} \ddd i_{N_{\tau}}\colon\, \tau^{[i_{1}]} = \cdots = \tau^{[i_{N_{\tau}}]} = \tau\right\}.
\end{equation}
The estimate of $\mathrm{DIC}_{\tau}$ can be computed using the posterior draws $\{(\boldsymbol{\theta}_{\tau}^{[i]},\, \tau^{[i]})\colon\, i \in \mathcal{I}_{\tau}\}$, where $\bar{D}_{\tau} \approx N_{\tau}^{-1} \sum_{i \in \mathcal{I}_{\tau}} D(\boldsymbol{\theta}_{\tau}^{[i]},\, \tau^{[i]})$ and $\bar{\boldsymbol{\theta}}_{\tau} \approx N_{\tau}^{-1} \sum_{i \in \mathcal{I}_{\tau}} \boldsymbol{\theta}_{\tau}^{[i]}$. The posterior model probability $p(\tau \vbar \mathbf{r})$ is approximated by $N_{\tau} / N$. For any $\tau$ such that $\mathcal{I}_{\tau} = \varnothing$, it can be assumed that $p(\tau \vbar \mathbf{r}) \approx 0$; thus such $\tau$ is excluded during computation. Alternatively, one can compute $\mathrm{DIC}_{\mathrm{ave}}$ by separately computing the averaged posterior mean of deviance
\begin{equation}
\label{eq:dbar_ave}
\bar{D}_{\mathrm{ave}} = \E_{\tau}(\bar{D}_{\tau} \vbar \mathbf{r}) = \E_{\tau}\left\{\E_{\boldsymbol{\theta}_{\tau}}[D(\boldsymbol{\theta}_{\tau},\, \tau) \vbar \tau,\, \mathbf{r}] \Bvbar \mathbf{r}\right\} = \E_{\boldsymbol{\theta}_{\tau},\, \tau}[D(\boldsymbol{\theta}_{\tau},\, \tau) \vbar \mathbf{r}],
\end{equation}
and the averaged effective number of parameters
\begin{equation}
\label{eq:pd}
p_{D}^{\mathrm{ave}} = \E_{\tau}(p_{D}^{\tau} \vbar \mathbf{r}) = \bar{D}_{\mathrm{ave}} - \sum_{\tau=1}^{M} D(\bar{\boldsymbol{\theta}}_{\tau},\, \tau) p(\tau \vbar \mathbf{r}).
\end{equation}
The averaged DIC is then given by $\mathrm{DIC}_{\mathrm{ave}} = \bar{D}_{\mathrm{ave}} + p_{D}^{\mathrm{ave}}$. Similarly, both terms can be approximated using a sample of posterior draws. For example, $\bar{D}_{\mathrm{ave}} \approx N^{-1} \sum_{i=1}^{N} D(\boldsymbol{\theta}_{\tau}^{[i]},\, \tau^{[i]})$.

In Table~\ref{tab:dic}, for each parametric model, we report the estimates of the single model DIC, the posterior mean of deviance, and the effective number of parameters; for the SP-GARCH model, we report the estimates of the averaged DIC, the averaged posterior mean of deviance, and the averaged effective number of parameters. For the table and the rest of the section, the notations $\mathrm{DIC}$, $\bar{D}$, and $p_{D}$ are overloaded to mean either the single model or the averaged version of the measures, depending on the model in question.

For all the stock indices, the DIC ranks SP-GARCH as the best model, followed by (in the order of increasing DIC) Beta-t, GJR, and GARCH models. Both measures $\bar{D}$ and DIC agree on the ranking of the models, indicating that the differences in model fit dominate the differences in complexity when ranked by the DIC. For all the models that allow $g(\cdot)$ to be asymmetric, the DIC estimates are exceedingly lower than those for the GARCH model. This supports the well known finding that it is important to model asymmetry for equity returns.

\begin{figure}[h]
\centering
\includegraphics[width = 1.0 \textwidth]{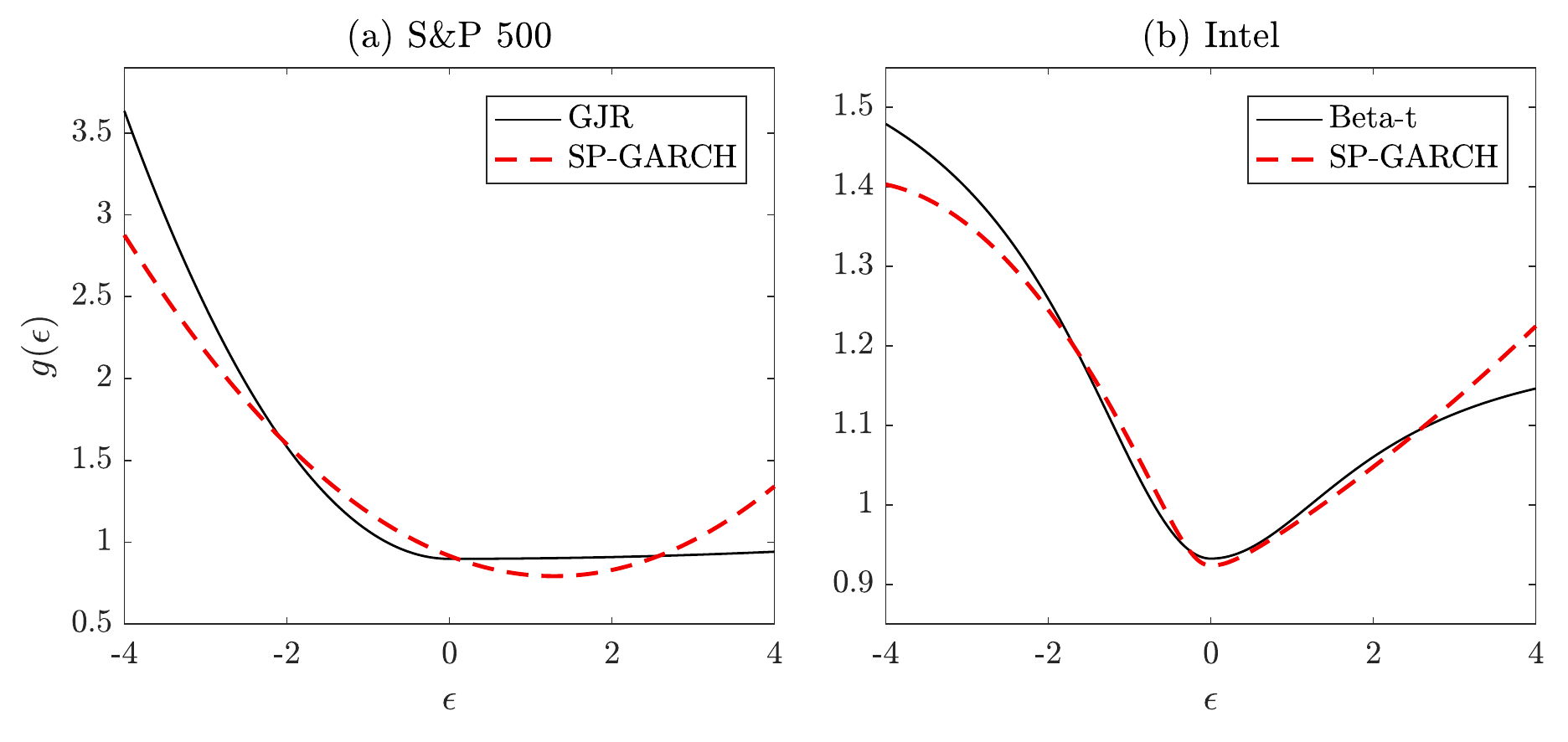}
\caption{(a) Posterior mean estimates of the coefficient functions of the GJR and SP-GARCH models for the S\&P 500 returns. (b) Posterior mean estimates of the coefficient functions of the Beta-t and SP-GARCH models are for the Intel returns.}
\label{fig:gfunc_comp}
\end{figure}

The fact that the SP-GARCH model stands out as the best amongst the asymmetric models calls for further investigation. Recall that, in Figure~\ref{fig:gfunc_emp}, the coefficient function estimates (a) -- (e) are minimised at points between 0 and 2, and that, in Table~\ref{tab:post_km}, most of the marginal posterior mass is on $K_{\mathbf{m}} = 0$ for the indices. In other words, the most adequate model for the stock indices is obtained when $g(\cdot)$ is a simple shifted quadratic function. This leads to the conclusion that, when modelling for stock indices, asymmetry is perhaps best modelled by allowing the minimum of $g(\cdot)$ to shift away from zero, as opposed to by allowing each side of zero to scale differently. To contrast the coefficient functions estimated under these two approaches, Figure~\ref{fig:gfunc_comp}(a) shows the posterior mean estimates of $g(\cdot)$ of both the GJR and SP-GARCH models for the S\&P 500 returns.

For the individual stocks, both the DIC and $\bar{D}$ estimates for the Beta-t and SP-GARCH models are substantially lower than those for the GARCH and GJR models. Unlike the index case, the DIC and $\bar{D}$ estimates for the GJR model are very similar to those for the GARCH model. This suggests either that asymmetry is of less importance for the chosen stocks, or that the type of asymmetry present in these returns can not be adequately captured by the GJR model.

For Apple, ARM, and Intel, the estimates of $\bar{D}$ for Beta-t are almost identical to those for SP-GARCH, while Beta-t achieves slightly lower DIC estimates. For Nvidia and SanDisk, SP-GARCH is ranked ahead of Beta-t by both DIC and $\bar{D}$.

By comparing the plots of the estimated coefficient functions of the SP-GARCH model to those of the Beta-t model, we find that the estimates given by the two models are very similar in their shapes. An example of such comparison is shown in Figure~\ref{fig:gfunc_comp}(b) for the Intel returns. Notice that the two functions are almost identical for $\epsilon_{t-1} \in [-2,\, 2]$, where most of the observations are found. Moreover, for large negative innovations, the coefficient function of the SP-GARCH shows a similar ``damping" behaviour to that of the Beta-t model. This finding provides evidence, at least for the chosen stocks, supporting the empirical validity of the Beta-t model.

The BMA approach allows SP-GARCH to adapt its complexity according to the complexity of the data. This is evident in the estimated effective number of parameters $p_{D}$. For the indices, the $p_{D}$ estimates for SP-GARCH are all close to 6, which are similar to those for the GJR and Beta-t models; for the individual stocks, the $p_{D}$ estimates for SP-GARCH are increased slightly to approximately 7, indicating that more complexity is needed to accommodate the more complex dynamics of the data.

\begin{table}[h]
\small
\hspace{-2.2cm}
\begin{tabular}{clcccccccccc}
\toprule
 &  & SP500 & FTSE & DAX & Nikkei & HSI & Apple & ARM & Intel & Nvidia & SanDisk \\
\midrule
\multirow{4}{*}{$\mathrm{DIC}$} & GARCH & 14205.0 & 14341.5 & 16785.8 & 16847.0 & 16914.8 & 23185.3 & 21852.1 & 21492.7 & 21717.0 & 26752.5 \\
 & GJR & 14037.0 & 14200.8 & 16682.0 & 16779.7 & 16856.0 & 23184.3 & 21852.6 & 21487.8 & 21709.1 & 26754.4 \\
 & Beta-t & 14015.1 & 14190.1 & 16667.5 & 16769.0 & 16838.6 & 23113.8 & 21773.4 & 21427.4 & 21648.3 & 26632.8 \\
 & SP-GARCH & 13977.6 & 14149.8 & 16630.9 & 16735.5 & 16827.4 & 23115.6 & 21776.1 & 21428.2 & 21637.7 & 26626.5 \\
\midrule
\multirow{4}{*}{$\bar{D}$} & GARCH & 14200.3 & 14336.7 & 16781.1 & 16842.0 & 16910.2 & 23180.9 & 21847.6 & 21488.2 & 21712.7 & 26748.4 \\
 & GJR & 14032.0 & 14195.6 & 16676.1 & 16773.8 & 16850.1 & 23179.1 & 21847.1 & 21482.3 & 21703.9 & 26749.3 \\
 & Beta-t & 14010.1 & 14184.9 & 16661.6 & 16763.1 & 16832.7 & 23108.1 & 21767.8 & 21421.7 & 21643.0 & 26627.4 \\
 & SP-GARCH & 13971.7 & 14144.0 & 16624.9 & 16729.3 & 16821.5 & 23108.3 & 21769.3 & 21420.9 & 21630.7 & 26619.4 \\
\midrule
\multirow{4}{*}{$p_{D}$} & GARCH & 4.8 & 4.8 & 4.7 & 5.0 & 4.6 & 4.4 & 4.6 & 4.5 & 4.3 & 4.1 \\
 & GJR & 5.0 & 5.2 & 5.9 & 5.9 & 5.9 & 5.2 & 5.5 & 5.5 & 5.1 & 5.1 \\
 & Beta-t & 5.0 & 5.2 & 5.9 & 5.9 & 5.9 & 5.6 & 5.5 & 5.7 & 5.3 & 5.4 \\
 & SP-GARCH & 5.9 & 5.9 & 6.0 & 6.2 & 5.9 & 7.2 & 6.8 & 7.3 & 7.0 & 7.2 \\
\bottomrule
\end{tabular}
\caption{Estimates of $\mathrm{DIC}$, $\bar{D}$, and $p_{D}$ for the stock indices and individual stocks. Some data labels are shortened to save space, i.e., SP500 = S\&P 500, and HSI = Hang Seng.}
\label{tab:dic}
\end{table}
\FloatBarrier

\FloatBarrier

\section{Conclusion}
\label{sec:conclusion}
In its original form, the volatility function governing the dynamics of the GARCH model has a very restricted form. However, financial markets may be subjected to dramatic changes, calling for flexible models that can adapt to the changes. The main contribution of this paper is to propose the SP-GARCH model, where the sequence of conditional variances $\{\sigma^{2}_{t}\}$ follows a functional coefficient autoregressive (FAR) model. Flexibility is achieved by allowing the coefficient function to be an arbitrary smooth function of the the lagged innovation. This smooth function is then approximated using regression spline basis functions based on a certain knot configuration. A strong feature of the proposed model is that the GARCH model and many of its popular extensions are special cases of the SP-GARCH model, differentiated by only the shape of the coefficient function $g(\cdot)$; inference on $g(\cdot)$ allows implicit testing of the simpler specifications. To perform inference, while accounting for the uncertainty in knot configuration, an approach based on Bayesian model averaging (BMA) is employed. To implement the BMA based approach, a carefully designed MCMC algorithm is developed to sample from the joint posterior distribution over the knot configurations and the parameter space. As illustrated by the trace plots, the sampler is able to move rapidly through the space on which the joint posterior distribution is defined. A simulation study shows that, when flexibility is required, the SP-GARCH model is, indeed, able to learn the shape of the coefficient function from the data. On the other hand, when the true data generating process is simple (namely, the original GARCH model), the SP-GARCH model, as a conditional volatility estimator, is as efficient as the simpler parametric models. Applications to ten real financial time series reveal that the coefficient functions for the stock indices are different in structural features from those for the individual stocks. A model comparison study based on the DIC confirms that the SP-GARCH model is, indeed, the most adequate description of the data. Furthermore, the SP-GARCH model is able to adapt its complexity according to the complexity of the data; when applied to real return series, the SP-GARCH model is often as parsimonious as the popular parametric models, which makes it very interpretable.

The SP-GARCH model can be naturally extended to incorporate realised measures of volatility constructed using high-frequency intra-daily data, such as realised variance, realised kernel, and realised range (See, for example, \citet{AndersenEtAl2001}, \citet{Barndorff-NielsenShephard2002}, \citet{BarndorffNielsenEtAl2008}, and \citet{ChristensenPodolskij2007}). For example, the volatility function of the SP-Realised-GARCH model can be written as $\sigma^{2}_{t} = \omega + g(x_{t-1}) \sigma^{2}_{t-1}$, where $x_{t-1}$ is a lagged realised measure.

\ifnotblinded
\section{Acknowledgement}
\label{sec:acknowledgement}
We thank Gareth W. Peters and Chris J. Oates for their comments on the manuscript. WYC was supported by the Australian Research Council Centre of Excellence for Mathematical and Statistical Frontiers (ACEMS).
\fi

\bibliographystyle{chicago}
\bibliography{spline}
\end{onehalfspace}
\end{document}